\documentclass[fleqn,usenatbib]{mnras}

\usepackage{newtxtext,newtxmath}

\usepackage[T1]{fontenc}
\usepackage{ae,aecompl}

\usepackage{graphicx}
\usepackage{booktabs,caption}
\usepackage[flushleft]{threeparttable}
\usepackage{rotating}
\usepackage{tabularx}
\usepackage{hyperref}
\usepackage{easylist}
\usepackage{bm}
\usepackage{amssymb,amsmath}
\usepackage[normalem]{ulem}

\newcommand{\numax}{\mbox{$\nu_{\rm max}$}}

\newcommand{\muHz}{\mbox{$\mu$Hz}}

\newcommand{\target}{OGLE-2017-BLG-1186}

\title[OGLE-2017-BLG-1186: asteroseismology and GPs]{OGLE-2017-BLG-1186: first application of asteroseismology and Gaussian processes to microlensing}

\author[S.-S.~Li et al.]{
\parbox{\textwidth}{
S.-S.~Li,$^{1,2}$\thanks{E-mail: lshuns@nao.cas.cn}
W.~Zang,$^{3,B,C,E}$
A.~Udalski,$^{4,A}$
Y.~Shvartzvald,$^{5,B,C,E}$
D.~Huber,$^{6}$
C.-U.~Lee,$^{7,8,C}$
T.~Sumi,$^{9,D}$
A.~Gould,$^{7,10,11,B,C}$
S.~Mao,$^{3,1}$
P.~Fouqu\'e,$^{12,13}$
% ...
\\[20pt]{\it \small Affiliations appear at the end of the paper}
}
}

\date{Accepted 2019 July 3. Received 2019 July 1; in original form 2019 April 16}

\pubyear{2019}

\begin{document}
\label{firstpage}
\pagerange{\pageref{firstpage}--\pageref{lastpage}}
\maketitle

\clearpage

\begin{abstract}
We present the analysis of the event OGLE-2017-BLG-1186 from the 2017 \emph{Spitzer} microlensing campaign. This is a remarkable microlensing event because its source is photometrically bright and variable, which makes it possible to perform an asteroseismic analysis using ground-based data. We find that the source star is an oscillating red giant with average timescale of $\sim 9$~d. The asteroseismic analysis also provides us source properties including the source angular size ($\sim 27~\mu{\rm as}$) and distance ($\sim 11.5$~kpc), which are essential for inferring the properties of the lens. When fitting the light curve, we test the feasibility of Gaussian Processes (GPs) in handling the correlated noise caused by the variable source. We find that the parameters from the GP model are generally more loosely constrained than those from the traditional $\chi^2$ minimization method. We note that this event is the first microlensing system for which asteroseismology and GPs have been used to account for the variable source. With both finite-source effect and microlens parallax measured, we find that the lens is likely a $\sim 0.045~M_{\odot}$ brown dwarf at distance $\sim 9.0$~kpc, or a $\sim 0.073~M_{\odot}$ ultracool dwarf at distance $\sim 9.8$~kpc. Combining the estimated lens properties with a Bayesian analysis using a Galactic model, we find a $\sim 35$ per cent probability for the lens to be a bulge object and $\sim 65$ per cent to be a background disc object.
\end{abstract}




\section{Introduction} \label{sec:intro}

Intrinsic source properties such as angular size and distance are crucial for the interpretation of lens physical properties. When a measured source angular radius $\theta_{\ast}$ is combined with the scaled source radius $\rho_{\ast}$, which is derived from fitting a light curve exhibiting finite-source effects \citep{Witt1994ApJ,Gould1994ApJ2,Nemiroff1994ApJ,Yoo2004ApJ,Choi2012ApJ}, we can obtain the angular Einstein radius $\theta_{\rm E}$ of the lens as
\begin{equation}\label{equ:thetae}
    \theta_{\rm E} = \frac{\theta_{\ast}}{\rho_{\ast}}~.
\end{equation}\label{thetae}

This, when combined with the microlens parallax $\pi_{\rm E}$, leads to an unambiguous mass measurement of the lens \citep{Gould1992ApJ}
\begin{equation}\label{equ:m}
M_{L} = \frac{\theta_{\rm E}}{\kappa\pi_{\rm E}}~,
\end{equation}
where $\kappa\equiv 4G/(c^2~{\rm au})\simeq 8.14~{\rm mas}/M_{\odot}$. In addition, if the distance to the source star $D_s$ is also determined, the lens distance $D_{L}$ can be derived by
\begin{equation}\label{equ:d}
D_{L}=\frac{\rm au}{\pi_{\rm E}\theta_{\rm E}+\pi_{s}}~,
\end{equation}
where $\pi_s\equiv {\rm au}/D_s$ is the parallax of the source~\citep{Gould1992ApJ,Gould2000ApJ}.

Generally, the source angular radius can be derived from the source's de-reddened colour and magnitude using colour/surface-brightness (CSB) relations~\citep[see, e.g.,][]{Kervella2004AA,Kervella2008AA,Boyajian2014AJ}, which can in turn be obtained by comparing the source position with the centroid of the `clump' of red giants on a colour-magnitude diagram (CMD)~\citep{Albrow2000ApJ,Yoo2004ApJ}. The basic assumption behind this method is that the source and the red clump experience the same extinction. This is reasonable for the majority of microlensing events because the vast majority of sources lie either in the Galactic bulge (which also contains the overwhelming majority of clump stars) or in the foreground disc but beyond most of the obscuring dust.  The latter typically occurs for disc lenses simply because the stellar scale height is several times larger than the dust scale height and typical sight lines intersect the bulge well above (or below) the dust scale height. Nevertheless, it warrants caution when dealing with events located near the Galactic plane~\citep{Bennett2012ApJ,Mroz2017AJ,Shvartzvald2018ApJ,Bennett2018AJ,Ranc2018arXiv}.

For low-latitude events, the lines of sight stay much closer to the Galactic plane, so the Galactic thin disc population (foreground and background) can have a significantly higher contribution to the microlens sources and lenses. Furthermore, the dust clouds can cause large extinction variations along the line of sight. All these anomalies can make the traditional CMD method unsuitable or cumbersome, leading to ambiguities in the source distance and angular radius.

There are several examples of previous microlenses at low latitudes that have ambiguous source distances. \cite{Street2016ApJ} found that the source in OGLE-2016-BLG-0966 was ambiguous between the foreground disc and bulge populations, leading to uncertainties in the derived properties of the lens and its planet. Subsequent analysis of the spectrum of the source supports the conclusion that it is in the bulge but was unable to completely resolve this degeneracy \citep{Johnson2017PASP}. As another example, \cite{Bennett2018AJ} reported a source star with an unusually red colour in the planetary event MOA-2011-BLG-291. In this case, the traditional assumption of a bulge source would yield a planetary system with $D_L\sim 7$~kpc. However, a more careful analysis that incorporated constraints on the distance to the source preferred a system with both the lens and source located in the foreground Galactic disc ($D_L\sim 4$~kpc). Likewise, \cite{Shvartzvald2018ApJ} found the source in UKIRT-2017-BLG-001 is inconsistent with the standard assumption that it is at the distance to the red clump. Rather, they found it is more consistent with being part of the far disc population. Uncertainties in the distances to the sources propagate to uncertainties in the distances to the lenses. Thus, constraining the source distance is important as it could affect the statistical study of the planet formation in different stellar environments (Galactic bulge versus disc) \citep{Calchi2015ApJ,Penny2016ApJ,Zhu2017AJ}.

Asteroseismology provides an alternative for deriving precise stellar properties~\citep[see, e.g.,][]{Brown1994ARAA,Christensen2004SoPh,Aerts2010aste.book}. In the most basic form, it is based on two global asteroseismic parameters, the frequency of maximum power $\nu_{\rm max}$ and the large-frequency separation $\Delta\nu$, which are approximately related to the stellar mass $M$ and radius $R$ as~\citep{Ulrich1986ApJ,Brown1991ApJ,Kjeldsen1995AA}
\begin{equation}\label{equ:astero1}
\Delta\nu \simeq \frac{(M/M_{\sun})^{1/2}}{(R/R_{\sun})^{3/2}}\Delta\nu_{\sun}~,
\end{equation}
\begin{equation}\label{equ:astero2}
\nu_{\rm max} \simeq \frac{M/M_{\sun}}{(R/R_{\sun})^{2}\sqrt{T_{\rm eff}/T_{\rm eff,\sun}}}\nu_{\rm max,\sun}~,
\end{equation}
where $T_{\rm eff}$ is the effective temperature, and the subscript `$\sun$' indicates parameters for the Sun. By combining these scaling relations with corresponding photometry and an extinction law, we can derive intrinsic source properties as well as distances either through the `direct method' or by `grid modelling' as described in \citet{Huber2017ApJ}.

Over the past decade, asteroseismology has become a powerful method to characterize host stars in transiting exoplanet systems~\citep{Kjeldsen2009IAUS,Christensen2010ApJ,Carter2012Sci,Huber2013ApJ,Huber2013Sci,Grunblatt2016AJ}. This has primarily benefited from the exquisite photometric performance of the \emph{Kepler} Mission~\citep{Stello2009ApJ,Kjeldsen2010AN,Gilliand2010PASP}. \emph{Kepler} detected solar-like oscillations in more than $500$ main-sequence and subgiant stars \citep{Chaplin2014ApJS}, and high-quality asteroseismic data are available on nearly $20,000$ red giants~\citep{Yu2018ApJS,Hon2019MNRAS}.
Nevertheless, synergy between the fields of microlensing and asteroseismology has not yet developed, in part because asteroseismic analysis requires long, continuous high-precision time-series photometry, which is hard to secure by current microlensing surveys. Fortunately, this might be revolutionized by the proposed \emph{Wide Field InfraRed Survey Telescope} (\emph{WFIRST},~\citealt{Spergel2015arXiv,Penny2018arXiv}). Owing to the high-precision astrometry and large aperture, \emph{WFIRST} is expected to yield roughly 1 million detections of oscillations in stars toward the Galactic bulge~\citep{Gould2015JKAS}. Furthermore, the candidate fields for the \emph{WFIRST} microlensing survey are near to the Galactic plane (see Figure 7 in \citealt{Penny2018arXiv} for the provisional fields for different \emph{WFIRST} designs), i.e., the aforementioned area where the traditional CMD method meets difficulties. Hence, as in the field of transiting exoplanets, the application of asteroseismology in the microlensing field might be productive in the relatively near future when we enter the era of \emph{WFIRST}.

Here we conduct the first asteroseismic analysis to a microlensing event. The event is \target, which has a low Galactic latitude, $b\simeq-1.8\deg$, lying in the candidate latitude region of the \emph{WFIRST} footprint ($-2.0\lesssim b\lesssim -0.5$). It has a very bright source star ($I \simeq 14.0$), which makes it possible to extract frequency information from ground-based observations. We use the $5$-yr OGLE-IV baseline data to perform asteroseismic measurements.

In spite of the benefits variable sources bring to the measurement of source properties, they become nuisances in the process of light curve modelling. The stellar variations show themselves in the microlensing modelling as correlated noise, which demands a proper treatment in order to reach the optimal fitting. Here we test a well-established technique called Gaussian processes (GPs)~\citep{Rasmussen2006} to tackle this correlated noise. GPs have been widely adapted to other exoplanet observations, including transit timing analysis~\citep{Gibson2012MNRAS,Evans2015MNRAS,Grunblatt2017AJ} and radial velocity measurements~\citep{Brewer2009MNRAS,Barclay2015ApJ,Grunblatt2016IAUFM,Czekala2017ApJ}, but they have not yet been specialized for application to microlensing. Our experiment proves the ability of GP model to tackle correlated noises in the micorlensing modelling. While the fitting results are not identical to those from the traditional $\chi^2$ minimization method, they are all consistent with each other within $\lesssim 3\sigma$.

The remainder of this paper is organized as follows. In Section~\ref{sec:obs}, we summarize the observations of \target. In Section~\ref{sec:analysis}, we present our methodology for light curve fitting. Two strategies are conducted: the traditional $\chi^2$ minimization method (Section~\ref{subsec:trad}) and the new GP method (Section~\ref{subsec:GP}).
The intrinsic source properties are derived using asteroseismic analysis in Section~\ref{sec:source}. And we interpret the physical properties of the lens in Section~\ref{sec:lens}. In Section~\ref{sec:dis}, we discuss our results and draw conclusions.

\section{Observations} \label{sec:obs}
OGLE-2017-BLG-1186 was first alerted by the Optical Gravitational Lensing Experiment (OGLE) collaboration on 2017 June 28 using its 1.3~m Warsaw Telescope equipped with a 1.4~${\rm deg}^2$ FOV mosaic CCD camera at the Las Campanas Observatory in Chile \citep{OGLEIV}. The event was located at equatorial coordinates $(\alpha, \delta)_{\rm J2000}$ = (17:58:46.95, $-27$:39:03.9), corresponding to Galactic coordinates $(\ell,b)=(2.58, -1.84)$. It lies in the OGLE field BLG504, which was observed with a cadence 3--10 observations per night. This event was also identified by the Microlensing Observations in Astrophysics (MOA) group as MOA-2017-BLG-396 on 2017 July 19th \citep{Bond2001}. The MOA group conducts a high-cadence survey toward the Galactic bulge using its 1.8~m telescope equipped with a 2.2~${\rm deg}^2$ FOV camera at the Mt.~John University Observatory in New Zealand \citep{MOA2016}. For this event, the cadence of the MOA observations is about $\Gamma = 3~{\rm hr}^{-1}$. The Korea Microlensing Telescope Network (KMTNet) group also observed this event, which it independently discovered as KMT-2017-BLG-0357 \citep{Kim2018a}, using its three 1.6~m telescopes equipped with 4~${\rm deg}^2$ FOV cameras at the Cerro Tololo International Observatory (CTIO) in Chile (KMTC), the South African Astronomical Observatory (SAAO) in South Africa (KMTS), and the Siding Spring Observatory (SSO) in Australia (KMTA) \citep{KMT2016} with a cadence of $\Gamma = 4~{\rm hr}^{-1}$. The vast majority of OGLE and KMTNet observations were carried out in the $I$-band, while MOA images were taken in a customized MOA-Red filter, which is similar to the sum of the standard Cousins $R$- and $I$-band filters. These surveys all had occasional $V$-band observations made solely to determine source colours. In addition, KMT data over the peak ($I < 12$ for KMTA and KMTS, $I < 12.5$ for KMTC) were excluded from the analysis due to problems caused by saturation.

OGLE-2017-BLG-1186 was initially selected as a `secret' target for the 2017 July 3 target upload to \emph{Spitzer} spacecraft.\footnote{The \emph{Spitzer} observation is a part of a large program measuring the Galactic distribution of planets in different stellar environments~\citep{Calchi2015ApJ,Zhu2017AJ}. Targets for \emph{Spitzer} observations can be selected `objectively' if they meet the specified objective criteria. Those events must be observed with a pre-specified cadence. Events that do not meet the criteria can still be chosen `subjectively' at any time for any reason,  but only data taken (or rather, made public) after this selection date can be used to calculate the planetary sensitivity of the events. In addition, events can be selected `secretly' without any announcement and become `subjectively' after the \emph{Spitzer} team makes a public announcement. See \citet{Yee2015ApJ.criteria} for a detailed description.} It was announced as a {\it Spitzer} target at UT 17:08 on 2017 July 6 prior to the first {\it Spitzer} observation because the event had a giant star source and thus it was recognized as a `Hollywood' event with high sensitivity to planets \citep{Hollywood1997}. Although it has no significance for the analysis presented in this paper, we note that the event met
the `objective' criteria for selection on 2017 July 17. In total, the {\it Spitzer} observations began on 2017 July 7 and ended on 2017 August 3 with a cadence of approximately 1 observation per 1.3 days; the `objective' observations began after HJD' $\sim 7956$.

Dense follow-up observations were taken after the {\it Spitzer} alert, with the aim of detecting and characterizing any planetary signatures. The follow-up teams include the Las Cumbres Observatory (LCO) global network, the Microlensing Follow-Up Network ($\mu$FUN, \citealt{mufun}), Microlensing Network for the Detection of Small Terrestrial Exoplanets (MiNDSTEp, \citealt{MINDSTEp}) and the University of Tasmania Greenhill Observatory. The LCO global network observed this event from its 1.0~m telescopes sited at CTIO, SAAO and SSO,  with the SDSS-$i'$ filter. The $\mu$FUN team provided observations from the 1.3~m SMARTS telescope at CTIO (CT13) with $V/I/H$-bands \citep{CT13}, the 0.4~m telescope at Auckland Observatory (Auckland) using a number 12 Wratten filter (which is similar to $R$-band), the 0.36~m telescope at Kumeu Observatory (Kumeu) in Auckland, the 0.36~m telescope at Turitea Observatory (Turitea) in the $R$-band, and the 0.36~m telescope at Possum Observatory (Pos) without a filter. Pos data were excluded from the analysis because they are flat over their two days of observations, giving no useful constraint on the model. The MiNDSTEp team monitored the events using the Danish 1.54~m telescope located at ESO's La Silla observatory in Chile, with a simultaneous two-colour instrument (wide visible and red; See Figure 1 of \citealt{Mind}). This event was also observed in the Bessell $I$-band by the 50-inch H127 telescope at the University of Tasmania (TAS)

Photometry of the OGLE, MOA, KMTNet, LCO, Auckland, Kumeu, Danish and TAS data were extracted using custom implementations of the difference image analysis \citep{Alard1998}: \citealt{Wozniak2000} (OGLE), \citealt{Bond2001} (MOA), \citealt{pysis} (KMTNet, LCO, Auckland, TAS and Kumeu) and \citealt{DanDIA} (Danish). The CT13, Pos and Turitea images were reduced using \texttt{DoPHOT}~\citep{dophot}. The {\it Spitzer} data were reduced using specialized software for crowded fields \citep{Spitzerdata}.

\section{Light curve analysis}\label{sec:analysis}

In this section, we present the process of light curve fitting. We first summarize the microlensing model adopted in Section~\ref{subsec:model}. Then we introduce our two fitting methods, the traditional $\chi^2$ minimization method and the Gaussian Processes in Section~\ref{subsec:trad} and Section~\ref{subsec:GP}, respectively. A brief description of our error rescaling strategy is also provided in Section~\ref{subsec:error}.

\subsection{Microlensing Model}\label{subsec:model}

The light curve of \target\, exhibits a standard symmetric Paczy\'nski curve~\citep{Paczynski1986ApJ} with clear finite-source effects shown in the peak (see Figure~\ref{fig:lc}). The event was intensively monitored by ground-based observatories and did not show any significant anomalies caused by multiple lenses or sources. Hence, we use a point lens with finite-source effects as our microlensing model. The formalism of this microlensing model can be found in~\citet{Yoo2004ApJ}.

The microlensing model with parallax and finite-source effects is described by six parameters ($t_0$, $u_0$, $t_{\rm E}$, $\rho_{\ast}$, $\pi_{\rm E,N}$, $\pi_{\rm E,E}$). Specifically, ($t_0$, $u_0$, $t_{\rm E}$) are the three Paczy\'nski parameters describing the light curve for a point lens with a point source \citep{Paczynski1986ApJ}: $t_0$ is the time of the maximum magnification, $u_0$ is the impact parameter (scaled to the angular Einstein radius $\theta_{\rm E}$), and $t_{\rm E}$ is the Einstein radius crossing time, with all parameters specified as being seen from Earth. $\rho_{\ast}=\theta_{\ast}/\theta_{\rm E}$ is the scaled source radius associated with the finite-source effects, where $\theta_{\ast}$ is the source angular radius. Finally, ($\pi_{\rm E,N}$, $\pi_{\rm E,E}$) are the north and east components of the microlens parallax vector, respectively.

Furthermore, we also introduce two flux parameters ($f_{s,n}$, $f_{b,n}$) on account of the possible blending effect for each observatory. Specifically, the observed flux for each dataset is modelled as $F_{{\rm lens}, n}(t)=A_n(t)f_{s,n}+f_{b,n}$, where $A_n(t)$ is the magnification at the $n$-th observatory as a function of time, which is characterized by the six microlensing parameters mentioned before. Due to the small separations between different ground-based sites compared with the projected Einstein radius, we approximate one magnification parameter $A_{\oplus} (t)$ for all the ground-based sites. That is, we ignore the so-called terrestrial parallax~\citep{Gould1997ApJ}. Nevertheless, the difference between $A_{Spitzer} (t)$ and $A_{\oplus} (t)$ is still significant.

This difference yields a measurement of the two-dimensional spaced-based microlens parallax~\citep{Refsdal1966MNRAS,Gould1994ApJ,Gould1995ApJ,Udalski2015ApJ}
\begin{equation}\label{equ:para}
\bm{\pi}_{\rm E} =\frac{\rm au}{D_{\perp}}\left(\Delta\tau,\Delta\beta\right)~,
\end{equation}
in which
\begin{equation}\label{equ:para2}
\Delta\tau\equiv\frac{t_{0,Spitzer}-t_{0,\rm\oplus}}{t_{\rm E}}~;~~\Delta\beta\equiv\pm u_{0,Spitzer}-\pm u_{0,\oplus}~,
\end{equation}
and ${D}_{\perp}$ is the projected distance between Earth and \emph{Spitzer}.

There are four possible values of $\bm{\pi}_{\rm E}$ resulting from the combination of different signs of $u_{0,Spitzer}$ and $u_{0,\oplus}$. These four values usually yield very similar light curve patterns, creating the well-known four-fold degeneracy~\citep{Refsdal1966MNRAS,Gould1994ApJ,Gould2004ApJ,Gould2013ApJ,Yee2015ApJ,Calchi2015ApJ}. The geometries for these four solutions can be found in Figure 2 of~\citet{Gould1994ApJ}. We specify the four solutions as $(+,+)$, $(+,-)$, $(-,+)$ and $(-,-)$, using the sign convention described in~\citet{Zhu2015ApJ}. Briefly, the first and second signs in each parenthesis indicate the signs of $u_{0,\oplus}$ and $u_{0,Spitzer}$, respectively.

Along with the finite-source effects, there is another proximity effect called limb darkening which is caused by the wavelength-dependent diminution of surface brightness from the centre of the disc to the limb of the star. As is customary, we adopt a linear limb-darkening law to consider the brightness profile of the source star~\citep{An2002ApJ}
\begin{equation}\label{equ:ld}
S_{\lambda}(\theta)=\bar{S}_{\lambda}\left[1-\Gamma_{\lambda}\left(1-\frac{3}{2}\cos\theta\right)\right]~,
\end{equation}
where $\bar{S}_{\lambda}\equiv f_{s,\lambda}/(\pi\theta_{\ast}^2)$ is the mean surface brightness of the source with $f_{s,\lambda}$ denoting the total source flux at wavelength $\lambda$, $\Gamma_{\lambda}$ is the limb-darkening coefficient at wavelength $\lambda$, and $\theta$ is the angular distance to the centre of the source. Based on the source properties derived in Section~\ref{sec:source}, assuming effective temperature $T_{\rm eff}\approx 3750$~K, surface gravity $\log g\approx 1$, microturbulent velocity $2$~km~s$^{-1}$, and metallicity $\log[M/H]=0$, we adopt $\Gamma_{R}=0.75$, $\Gamma_{I}=0.59$, $\Gamma_{H}=0.36$, and $\Gamma_{3.6\mu\rm{m}}=0.20$ from~\citet{Claret2011AA}. For MOA data, we estimate $\Gamma_{R'}$ as $(\Gamma_{I}+\Gamma_{R})/2$. For simplicity, we use $\Gamma_I$ for LCO SDSS-$i'$ data. The $\Gamma_{\rm Danish}$ is fitted as a free parameter because of the non-standard filter.

We note that the limb-darkening effect is a high-order effect in microlensing modelling, hence measurement of its coefficients from light-curve analysis is usually a difficult task. In the cases where dense-coverage and good-quality data are available, the measured coefficients are generally in good agreement with theoretical values~\citep{Choi2012ApJ,Shvartzvald2018arXiv}. In other cases, large differences (but with large uncertainties) between the measured results and the theoretical values are common~\citep{Fouque2010AA}. For this event, fitting limb-darkening coefficients is even harder due to the modifications introduced by the variable source. We have tried to fit the limb-darkening coefficients but cannot converge on results. On the other hand, the source properties are well constrained for this event due to the asteroseismic analysis, therefore theoretical values are acceptable for this event.

\subsection{The traditional method with white noise assumption}\label{subsec:trad}

The traditional $\chi^2$ minimization method is based on an implicit assumption\footnote{It is straightforward to incorporate correlated errors into the $\chi^2$ formalism, e.g., \cite{2003astro.ph.10577G}, but in practice this is rarely done.} that the residuals (difference between the observed values and the predicted values) are independent for distinct times, i.e., that the noise is white. Then the best model is found by minimizing $\chi^2$ with the form
\begin{equation}\label{equ:ltrad}
\chi^2\equiv\sum^{N}_{i=1}\left(\frac{F_i-F_{{\rm lens},i}}{\sigma_i'}\right)^2~,
\end{equation}
where $\{F_i,~F_{{\rm{lens}},i}\}^{N}_{i=1}$ are data points (fluxes) from observations and microlens modelling, respectively, and $\{\sigma_i'\}^{N}_{i=1}$ are the rescaled photometric errors. Their connection to the observational errors is specified in Section~\ref{subsec:error}. If the errors are Gaussian, then the likelihood is given by $\mathcal{L}=\exp\left(-\chi^2/2\right)$, so that minimizing $\chi^2$ is equivalent to maximizing $\mathcal{L}$. We then estimate the microlensing parameters using a Markov Chain Monte Carlo (MCMC) analysis through the \texttt{emcee} ensemble sampler developed by \citet{Foreman2013PASP}. However, even if the errors in the data points are not Gaussian, the errors in the derived parameters will usually be Gaussian, as long as the number of data points is reasonably large (see \cite{2003astro.ph.10577G} for a more detailed discussion).

The best-fit parameters with $1\sigma$ uncertainties for the four-fold degenerate solutions are shown in Table~\ref{table:trad}. The $(+,-)$ solution is slightly preferred over the other ones, but all the solutions are degenerate within $\Delta\chi^2\lesssim5$. The best-fit model for OGLE data is shown in Figure~\ref{fig:lc} with the magenta dashed line. In addition, the source star was detected by the Two Micron All Sky Survey (2MASS) \citep{2MASS}, which shows $H = 10.715\pm0.034, K = 10.318\pm0.033$. By calibrating CT13 $H$-band photometry to the 2MASS photometric system, we obtain $H_{\rm s} = 10.753\pm0.030$ from the best-fit model, and OGLE $I$-band photometry has $\sim7.9\%$ blended light. Thus, the source is blended, and we adopt $H_{\rm s} = 10.75\pm0.03, K_{\rm s} = 10.36\pm0.03$ for future analysis.

\subsection{Gaussian-process modelling of the correlated noise}\label{subsec:GP}

\begin{figure*}
\centering
\resizebox{\hsize}{!}{\includegraphics{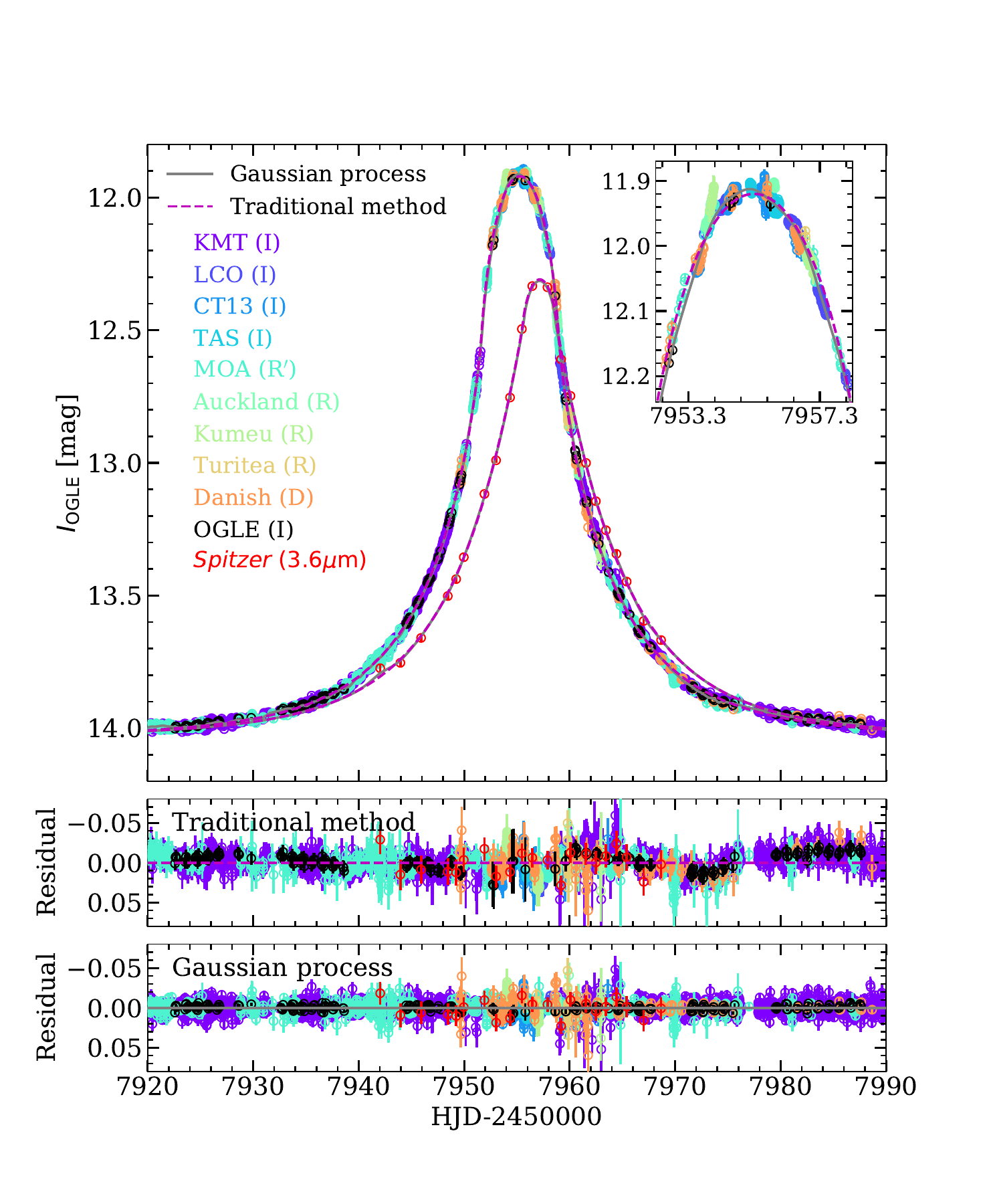}}
\caption{Top panel: the light curve of \target\ with the best-fit models for OGLE (or $I$-band) data, i.e., the $(+,-)$ case, from both the traditional method (magenta dashed line) and the GP method (gray solid line). The inset shows the peak in greater detail. Lower panels: residuals for each method.}
\label{fig:lc}
\end{figure*}

A noticeable trend remains in the residuals shown in Figure~\ref{fig:lc} using the traditional $\chi^2$ minimization method, which indicates a violation of the white noise assumption. Based on a detailed analysis in Section~\ref{sec:source}, this trend mainly results from the quasi-periodic variability of the source and is not associated with the lens. Obtaining an optimal fit requires some method to account for this correlated noise, for which we turn to Gaussian Processes (GPs).

The GP model is one of the most popular non-parametric models for regression problems in the machine learning community. A non-parametric model does not interpret the training data with a finite-dimensional parameter vector, instead it places a distribution over a (usually infinite) number of functions to interpret the data and makes predictions based on all the training data. This purely data-driven approach makes it flexible enough to handle stochastic behaviours of the data using only a few hyperparameters and without suffering inconsistency problems.  A more comprehensive introduction can be found in \citet{Rasmussen2006}.

The GP model allows us to handle the deterministic and stochastic components of the data with a general multivariate Gaussian distribution
\begin{equation}\label{equ:distri}
p(\bm{F}|\bm{t},\bm{\phi},\bm{\theta}) = \mathcal{N}(\bm{m}(\bm{t},\bm{\phi}),\bm{K}(\bm{t},\bm{\theta}))~,
\end{equation}
where $\bm{F}(\bm{t})$ is the set of observations with $\bm{F}$ and $\bm{t}$ denoting the vectors of fluxes and time, respectively. $\mathcal{N}$ indicates a Gaussian distribution with the mean function $\bm{m}(\bm{t},\bm{\phi})$ and covariance matrix $\bm{K}(\bm{t},\bm{\theta})$, where $\bm{\phi}$ is a vector of the microlensing parameters as described in Section~\ref{subsec:model}, and $\bm{\theta}$ is a vector of the hyperparameters characterizing the covariance matrix.\footnote{In the GP framework, both $\bm{\phi}$ and $\bm{\theta}$ are known as hyperparameters, because they are used to specify the distribution itself as opposed to any specific modelling functions. But here we refer to $\bm{\theta}$ as hyperparamters and keep $\bm{\phi}$ as the microlensing parameters in order to maintain a natural connection to the traditional method.}

The mean function $\bm{m}(\bm{t},\bm{\phi})$ controls the model's deterministic component, which in our case, is just the aforementioned microlensing model, while the covariance matrix $\bm{K}(\bm{t},\bm{\theta})$ is what GP models use to specify all the stochastic variations biased from the mean function. Each element of the covariance matrix is specified by a covariance function (aka kernel) $k(\bm{t},\bm{t}')$.

The kernel encapsulates the core of GPs. It defines nearness and similarity between data points. There are many kernels with different properties~\citep{Rasmussen2006}. For our purpose, we adopt the kernel~\citep{Foreman2017AJ}
\begin{equation}\label{equ:kernel}
k(\tau_{ij})=S_0\omega_0\exp\left(-\frac{1}{\sqrt{2}}\omega_0\tau_{ij}\right)\cos\left(\frac{\omega_0\tau_{ij}}{\sqrt{2}}-\frac{\pi}{4}\right)+\sigma_i'^2\delta_{ij}~,
\end{equation}
where $\{\sigma_{i}'^2\}^N_{i=1}$ are the rescaled errors, $\delta_{ij}$ is the Kronecker delta, and $\tau_{ij}=|t_i-t_j|$ specifies the time separation between data points. There are two hyperparameters associated with this kernel: $\omega_0$ determines the `closeness' between data points, and $S_0$ specifies the maximum amplitude of the covariance. The combination of an exponential term with a trigonometric term enables this kernel to handle quasi-periodic variations. Indeed, it has been widely adopted to model stellar granulation background in the literature of asteroseismic analysis~\citep{Harvey1985ESASP,Huber2009CoAst,Michel2009A,Kallinger2014A}.

We can now construct a distribution that can model the data with both microlensing effects and correlated noise being considered (Equation~\ref{equ:distri}). As the distribution is a multivariate Gaussian, the log-likelihood function is just
\begin{equation}\label{equ:gplogl}
\ln\mathcal{L}(\bm{\theta},\bm{\phi})=-\frac{1}{2}\bm{r}^{T}\bm{K}^{-1}\bm{r}-\frac{1}{2}\ln|\bm{K}|-\frac{N}{2}\ln(2\pi)~,
\end{equation}
where $\bm{r}=\bm{F}-\bm{m}$ is the vector of residuals from the mean function (microlensing model). Once we have the likelihood function, the microlensing parameters $\bm{\phi}$ and hyperparameters $\bm{\theta}$ can again be estimated via the MCMC analysis.

The key practical problem of GPs is that the computational cost scales as the cube of the number of data points due to the inverse and determinant of matrix $\bm{K}$ shown in Equation~(\ref{equ:gplogl}). The cubic scaling is prohibitive for large data sets. We here adopted the \texttt{celerite} algorithm developed by \citet{Foreman2017AJ} to perform the calculation. For one-dimensional data sets, this algorithm can compute the likelihood with the computational cost scaling linearly with the number of data points. This linear scaling is achieved by exploiting the semi-separable structure in a specific class of covariance matrices, specifically, matrices generated by a mixture of exponentials (see \citealt{Foreman2017AJ} for detailed discussions of the method and comparisons to other methods). To deal with the flux parameters $(f_{s,n},f_{b,n})$, we fix the baseline fluxes (i.e., $f_{s,n}+f_{b,n}$) as well as the source flux ratios $r_{s,n}$~($=f_{s,n}/f_{s,{\rm OGLE}}$) for each observatory using the best-fit results from the traditional method, and free $f_{b,{\rm OGLE}}$ in the chain.

\begin{figure*}
\centering
\resizebox{\hsize}{!}{\includegraphics{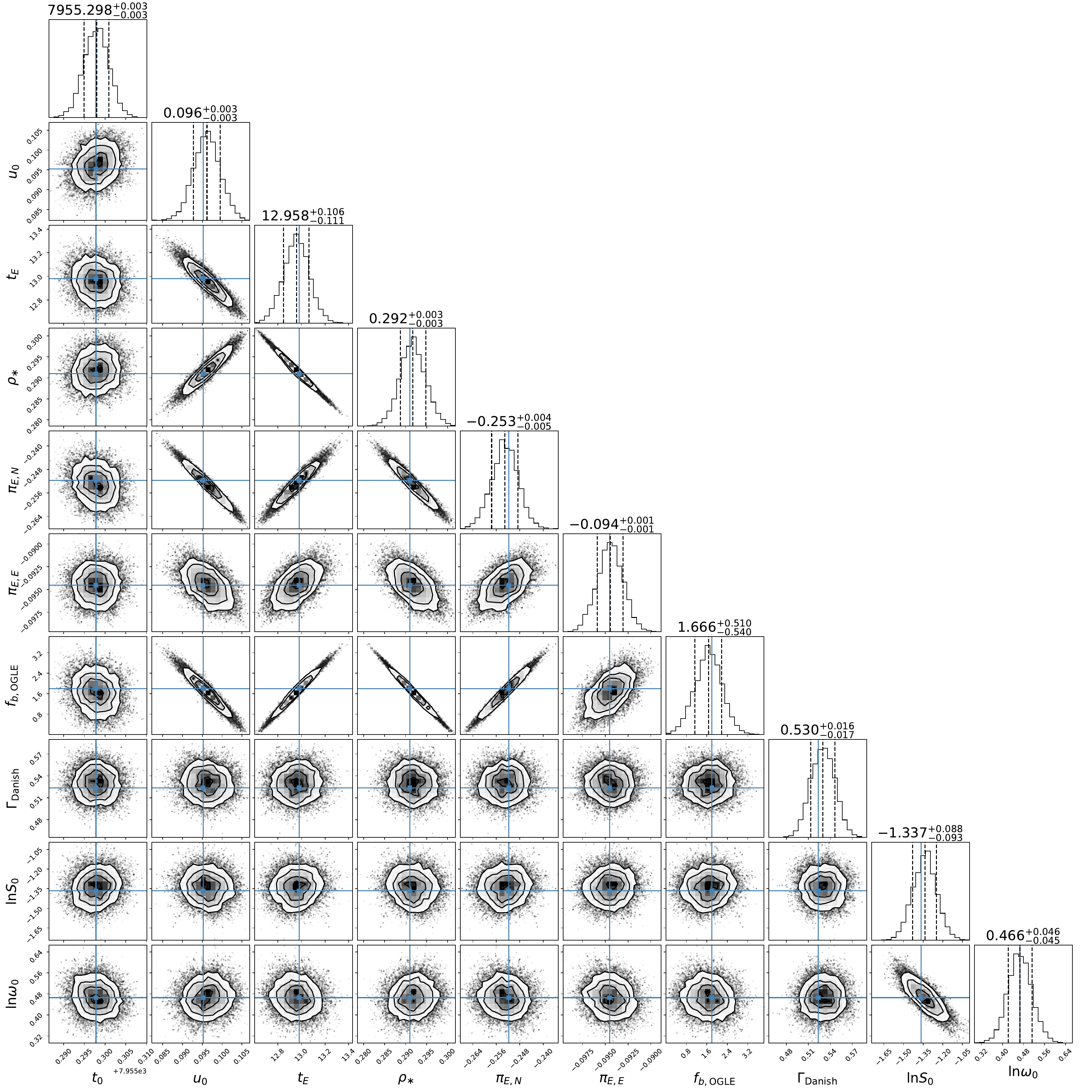}}
\caption{Parameter distributions for GP modelling. The first eight parameters are components of our microlensing model, and the last two parameters are hyperparameters of the kernel. The vertical dashed line indicates the median values and $1\sigma$ credible regions. The blue lines indicate the best fitted values (maximum likelihood).}
\label{fig:MCMC}
\end{figure*}

The new fitted light curve is shown as the solid black line in Figure~\ref{fig:lc}. The improvement of modelling is noticeable from comparing the two sets of residuals from traditional (middle panel) and GP (bottom panel) methods. These results lend some credibility to the GP model in handling correlated noises in microlensing signals. The posterior distributions for all the free parameters are shown in Figure~\ref{fig:MCMC}. As can be seen, all the parameters are well converged, and almost no degeneracy is shown between microlensing parameters and hyperparameters in this specific event.

Nevertheless, things become complex when estimating the microlensing parameters (see Table~\ref{table:GP}). First, the parameters derived from the GP model do not perfectly agree with those from the traditional $\chi^2$ method. For example, for the $(+,-)$ solution, the microlensing parameters $u_0$, $t_{\rm E}$, $\rho_{\ast}$ and $\pi_{{\rm E}, N}$ differ by $\gtrsim 3\sigma$, $\gtrsim 1\sigma$, $\gtrsim 1\sigma$ and $\gtrsim 2\sigma$, respectively. Although these levels of differences are likely from numerical uncertainties, they can also be the consequence of the degeneracy between the source oscillation period and the microlensing parameters. Second, the parameters derived from the GP model are generally more poorly constrained. This is reasonable since extra degrees of freedom usually introduce extra uncertainties. Chances are that the traditional method underestimated the uncertainties. In this aspect, the errors derived from the GP model are more realistic. Nevertheless, mainly due to the existence of blending effects, which are common to microlensing events but rare to other exoplanet observations (see e.g. \citealt{Grunblatt2017AJ,Czekala2017ApJ}), there are also some uncertainties associated with the GP model itself in microlensing modelling. Practical problems like how to properly deal with the different blending effects in different observations, how to perform error rescaling, still require better understanding. In addition, theoretical problems like the set of GP hyperparameters, the possible degeneracy between microlensing parameters and hyperparameters, demand more careful numerical experiments. Thoroughly solving all these issues is beyond the scope of this work. Therefore, in this event, we still adopt the microlensing parameters from the traditional method to derive the physical parameters, before we are fully confident about our GP model. Fortunately, due to the high magnification, the results derived from both modelling methods are generally consistent within $\lesssim 3\sigma$. In other words, the choice of modelling methods does not affect the final physical interpretation of the lens properties in this event.

\subsection{Error rescaling}\label{subsec:error}

The errors from photometric measurements typically overlook contributions from systematics, underestimating (or overestimating) true errors. Hence, the photometric errors are often renormalized in microlensing analyses~\citep{Yee2012ApJ,Street2013ApJ,Skowron2016AcA,Shin2018ApJ}.

For simplicity, we adopt the conventional strategy for both the traditional and GP methods. That is, regardless of how we perform our predictions, once we obtained the model predicted results, we construct the $\chi^2$ parameter by
\begin{equation}\label{equ:chi}
\chi^2 = \sum^{N}_{i=1}\left(\frac{M_i-M_{\rm{mod},i}}{\sigma_i'}\right)^2~,
\end{equation}
where $M_i$ and $M_{\rm{mod},i}$ are the $i$-th data points (magnitudes) from the real data and the modelling predictions, respectively. $\sigma_i'=k\sqrt{\sigma_i^2+e_{\rm min}^2}$ is the rescaled photometric error of the $i$-th data point, with $\sigma_i$ indicating the original error.

The values of $e_{\rm min}$ and $k$ are chosen such that the cumulative $\chi^2$ distribution for each set of data is approximately linear as a function of source magnification, and the total $\chi^2$ is equal to the number of points in that data set. In practice, we perform an MCMC analysis to find the best values of $e_{\rm min}$ and $k$ based on the criterion that $\sum^{N}_{i=1}\left(\chi^2_{i}-1\right)^2$ is minimized, where $\chi^2_{i}=\left(\left(M_i-M_{\rm{mod},i}\right)/\sigma_i'\right)^2$. The values of $k$ and $e_{\rm min}$ for the major data sets are listed in Table~\ref{table:error}. For the other data sets, we simply set $e_{\rm min}=0$ and adjust $k$ to enforce $\chi^2=N$. This is legitimate when the event is bright and the Poisson flux errors are small~\citep{Yee2012ApJ,Shin2018ApJ}. All the error rescaling processes are done based on the best-fit model,
i.e., the $(+,-)$ case.

We note that, in the case of the OGLE data, the photometric errors are updated using an empirical model provided by \citet{Skowron2016AcA}, before applying the aforementioned error rescaling process. Hence, the OGLE data are preprocessed such that $e_{\rm min}=0$.

\begin{table*}
\begin{threeparttable}
\caption{Best-fit parameters from the traditional method}
\label{table:trad}
\begin{tabular}{lrrrrrrrr}
\toprule
 &
\multicolumn{1}{c}{$(+,+)$} &
\multicolumn{1}{c}{$(+,-)^a$} &
\multicolumn{1}{c}{$(-,+)$} &
\multicolumn{1}{c}{$(-,-)$} \\
\midrule
\vspace{1ex}
$\chi^2 / N_{\rm data}$	&
11561.5 / 11577	&
11556.3 / 11577	&
11561.3 / 11577	&
11558.5 / 11577	\\
 \vspace{1ex}
$t_0$ (HJD')$^b$          &
$7955.2944^{+0.0014}_{-0.0013}$       &
$7955.2931^{+0.0013}_{-0.0014}$       &
$7955.2945^{+0.0014}_{-0.0014}$       &
$7955.2939^{+0.0014}_{-0.0013}$       \\
 \vspace{1ex}
$u_0$                     &
$0.0811^{+0.0016}_{-0.0016}$           &
$0.0811^{+0.0016}_{-0.0016}$           &
$-0.0813^{+0.0015}_{-0.0016}$           &
$-0.0806^{+0.0017}_{-0.0015}$          \\
 \vspace{1ex}
$t_{\rm E}$ (days)     &
$13.1326^{+0.0349}_{-0.0354}$           &
$13.1241^{+0.0350}_{-0.0353}$           &
$13.1233^{+0.0365}_{-0.0337}$           &
$13.1401^{+0.0350}_{-0.0349}$          \\
 \vspace{1ex}
$\rho_{\ast}$             &
$0.2868^{+0.0010}_{-0.0009}$           &
$0.2869^{+0.0010}_{-0.0009}$           &
$0.2869^{+0.0009}_{-0.0010}$           &
$0.2866^{+0.0010}_{-0.0009}$          \\
 \vspace{1ex}
$\pi_{{\rm E},{\it N}}$               &   $0.1186^{+0.0023}_{-0.0020}$           &
$-0.2371^{+0.0027}_{-0.0029}$           &
$0.2286^{+0.0026}_{-0.0025}$           &
$-0.1272^{+0.0023}_{-0.0025}$          \\
 \vspace{1ex}
$\pi_{{\rm E},{\it E}}$               & $-0.1045^{+0.0012}_{-0.0012}$          &
$-0.0941^{+0.0013}_{-0.0012}$          &
$-0.1111^{+0.0012}_{-0.0012}$          &
$-0.0958^{+0.0012}_{-0.0012}$         \\
 \vspace{1ex}
$\Gamma_{\rm Danish}$     &
$0.5153^{+0.0234}_{-0.0240}$           &
$0.5058^{+0.0241}_{-0.0242}$           &
$0.5062^{+0.0255}_{-0.0235}$           &
$0.5113^{+0.0263}_{-0.0242}$          \\
 \vspace{1ex}
$f_{s,{\rm OGLE}}$$^c$     &
$35.79^{+0.17}_{-0.17}$           &
$35.77^{+0.17}_{-0.17}$           &
$35.82^{+0.16}_{-0.18}$           &
$35.79^{+0.17}_{-0.17}$          \\
$f_{b,{\rm OGLE}}$     &
$2.82^{+0.17}_{-0.17}$           &
$2.85^{+0.17}_{-0.17}$           &
$2.80^{+0.17}_{-0.16}$           &
$2.83^{+0.17}_{-0.17}$          \\
\bottomrule
\end{tabular}
    \begin{tablenotes}
      \small
      \item$^a$ The solution on which error rescaling process based.
      \item $^b$ HJD' is HJD$-2450000$.
      \item $^c$ We adopt $I = 18$ as the magnitude zeropoint.
    \end{tablenotes}
  \end{threeparttable}
\end{table*}

\begin{table*}
\begin{threeparttable}
\caption{Best-fit parameters with the Gaussian processes}
\label{table:GP}
\begin{tabular}{lrrrrrrrr}
\toprule
 &
\multicolumn{1}{c}{$(+,+)$} &
\multicolumn{1}{c}{$(+,-)^a$} &
\multicolumn{1}{c}{$(-,+)$} &
\multicolumn{1}{c}{$(-,-)$}\\
\midrule
\vspace{1ex}
$\ln\mathcal{L}$	&
$-3563.8$	&
$-3559.7$ &
$-3565.6$	&
$-3561.1$	\\
 \vspace{1ex}
$t_0$ (HJD')$^b$          &
$7955.2994^{+0.0031}_{-0.0030}$       &
$7955.2978^{+0.0030}_{-0.0030}$       &
$7955.2971^{+0.0032}_{-0.0031}$       &
$7955.2968^{+0.0031}_{-0.0030}$       \\
 \vspace{1ex}
$u_0$                     &
$0.0964^{+0.0034}_{-0.0032}$           &
$0.0952^{+0.0034}_{-0.0034}$           &
$-0.0962^{+0.0036}_{-0.0037}$           &
$-0.0962^{+0.0035}_{-0.0035}$          \\
 \vspace{1ex}
$t_{\rm E}$ (days)     &
$12.9500^{+0.1083}_{-0.1072}$           &
$12.9807^{+0.1059}_{-0.1106}$           &
$12.9297^{+0.1236}_{-0.1191}$           &
$12.9752^{+0.1120}_{-0.1183}$          \\
 \vspace{1ex}
$\rho_{\ast}$             &
$0.2920^{+0.0031}_{-0.0030}$           &
$0.2910^{+0.0031}_{-0.0030}$           &
$0.2923^{+0.0034}_{-0.0034}$           &
$0.2914^{+0.0033}_{-0.0030}$          \\
 \vspace{1ex}
$\pi_{{\rm E},{\it N}}$               &   $0.1146^{+0.0010}_{-0.0009}$           &
$-0.2518^{+0.0045}_{-0.0045}$           &
$0.2461^{+0.0048}_{-0.0048}$           &
$-0.1222^{+0.0009}_{-0.0009}$          \\
 \vspace{1ex}
$\pi_{{\rm E},{\it E}}$               & $-0.1038^{+0.0014}_{-0.0013}$          &
$-0.0945^{+0.0014}_{-0.0014}$          &
$-0.1128^{+0.0016}_{-0.0016}$          &
$-0.0962^{+0.0014}_{-0.0015}$         \\
 \vspace{1ex}
$\Gamma_{\rm Danish}$     &
$0.5335^{+0.0161}_{-0.0163}$           &
$0.5235^{+0.0164}_{-0.0165}$           &
$0.5242^{+0.0157}_{-0.0156}$           &
$0.5255^{+0.0157}_{-0.0163}$          \\
 \vspace{1ex}
$f_{b,{\rm OGLE}}$$^c$     &
$1.63^{+0.50}_{-0.53}$           &
$1.80^{+0.51}_{-0.54}$           &
$1.59^{+0.58}_{-0.58}$           &
$1.70^{+0.53}_{-0.57}$          \\
 \vspace{1ex}
$\ln S_0$     &
$-1.3117^{+0.0919}_{-0.0869}$           &
$-1.3664^{+0.0875}_{-0.0934}$           &
$-1.3283^{+0.0950}_{-0.0890}$           &
$-1.3581^{+0.0916}_{-0.0867}$          \\
 \vspace{1ex}
$\ln \omega_0$     &
$0.4479^{+0.0487}_{-0.0444}$           &
$0.4659^{+0.0458}_{-0.0448}$           &
$0.4635^{+0.0478}_{-0.0461}$           &
$0.4743^{+0.0438}_{-0.0436}$          \\
\bottomrule
\end{tabular}
    \begin{tablenotes}
      \small
      \item$^a$ The solution on which error rescaling process based.
      \item $^b$ HJD' is HJD$-2450000$.
      \item $^c$ We adopt $I = 18$ as the magnitude zeropoint.
    \end{tablenotes}
  \end{threeparttable}
\end{table*}

\begin{table}

\caption{Error rescaling}
\label{table:error}
\begin{tabularx}{\columnwidth}{l @{\extracolsep{\fill}} ccc}
\toprule
Observatory & $e_{\rm min}$ &  \multicolumn{2}{c}{$k$} \\
\cline{3-4}
 &  & Traditional & GP \\
\midrule
OGLE &  $0$ & $2.32$ & $0.62$ \\
MOA &  $0.002$ & $4.74$ & $2.45$ \\
KMTA02 &  $0.007$ & $1.53$ & $0.61$ \\
KMTA42 &  $0.012$ & $0.91$ & $0.54$ \\
KMTC02 &  $0.014$ & $0.80$ & $0.49$ \\
KMTC42 &  $0.008$ & $1.12$ & $0.59$ \\
KMTS42 &  $0.006$ & $1.61$ & $0.86$ \\
\bottomrule
\end{tabularx}
\parbox{\textwidth}{\small
\vspace{1ex}
All values are related to errors in magnitudes.}
\end{table}

\section{The source: an oscillating red giant}\label{sec:source}

The source star of \target\ was detected as a long-period variable in the OGLE-III fields toward the Galactic bulge by \citet{Soszynski2013AcA}. More specifically, it was identified as OGLE-BLG-LPV-153890 and classified as an OGLE small amplitude red giant. The primary period found by their period-searching code is $T\sim7.7$~days. This period simply corresponds to the peak of a single frequency ($\nu\sim1.5~\muHz$), while the measurement of the global asteroseismic parameter \numax\ is more sophisticated. It includes processes of removing fine structures from individual modes and smoothing (see Section~\ref{subsec:astero}). These processes are important for solar-like oscillations, since the `peak frequency' for a given star is changing due to the stochastic excitation and damping. Therefore, it is not a surprise that the peak of a single frequency is somewhat different (within $1.5\sigma$) from the global asteroseismic parameter \numax.

This variable feature opens the possibility to obtain source properties through an asteroseismic analysis. In this section, we perform both asteroseismic analysis (Section~\ref{subsec:astero}) and CMD analysis (Section~\ref{subsec:CMD}) to characterize the source star. Asteroseismic analysis provides additional parameters like the distance to the source that cannot be obtained through traditional CMD analysis. On the other hand, CMD analysis can provide extinction information required for distance estimation. The angular radius obtained from both analyses can serve as a cross check.

\subsection{Asteroseismic analysis}\label{subsec:astero}

\begin{figure*}
\centering
\resizebox{\hsize}{!}{\includegraphics{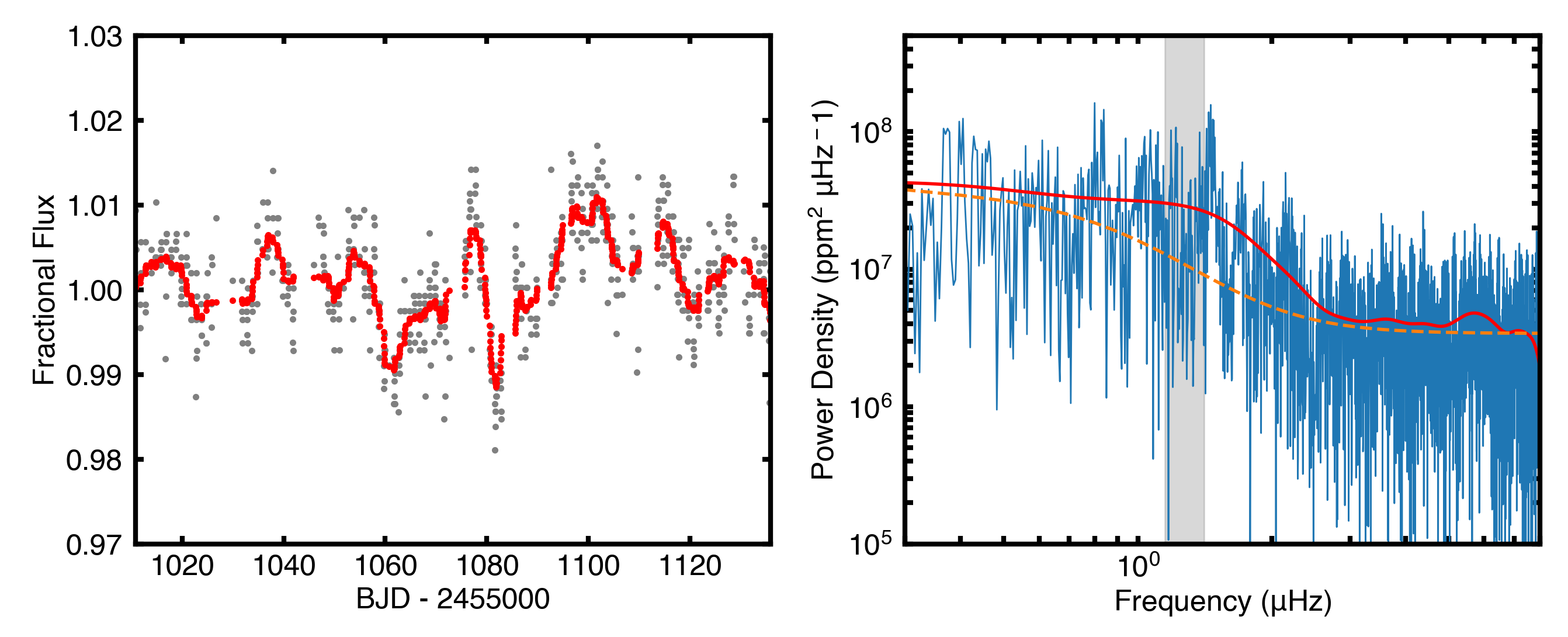}}
\caption{Left: Subset of the OGLE-IV baseline flux spanning 125~days. The red line shows a boxcar smoothing with a width of 0.5~days. The quasi-periodic variability with a timescale of 9~days is due to convection-driven oscillations. Right: power spectrum of the full light curve after removing the microlensing event. The orange dashed line shows the background model, and the red line is a heavily smoothed version of the power spectrum used to measure the frequency of maximum power. $\numax = 1.28 \pm 0.13\,\muHz$ is the peak of the smoothed curve after subtracting the background model (the shaded grey area shows the $1\sigma$ region of \numax).}
\label{fig:seismo}
\end{figure*}

Red giants exhibit solar-like oscillations driven by near-surface convection \citep{Hekker2017AApR}, reaching periods of weeks on the upper red-giant branch \citep{Huber2010CoAst,Mosser2011APP}. The power spectrum of \target\ after removing the microlensing event shows a typical correlated background noise due to stellar granulation \citep[e.g.,][]{Mathur2011ApJ}, superimposed with a Gaussian-shaped power excess due to oscillations (Figure \ref{fig:seismo}).

To measure global asteroseismic parameters we modelled the source variability in the Fourier domain using the methodology described in \citet{Huber2009CoAst}, yielding a frequency of maximum power of $\numax = 1.28 \pm 0.13\,\muHz$. The approximate amplitude per radial mode is $\sim800$~ppm in the $I$-band, consistent with the expected amplitude for red giants in this evolutionary stage \citep{Huber2011ApJ}.

The \numax\ is measured in the same way as it is measured for the Sun, since the scaling relation shown in Equation~(\ref{equ:astero2}) is based on some dependence of the stellar parameters on the observed solar values. For the method we are using, both \numax\ and $\nu_{\rm max,\sun}$ are measured by combining a given granulation model with a heavily smoothed power spectrum (a standard method originally suggested by \citealt{Kjeldsen2008ApJ}). As described in \citet{Kjeldsen2008ApJ}, the smoothing length is typically tied to some factors times the expected large-frequency separation. However, early M giants are not as well studied, since most Kepler stars have lower luminosity. Therefore, there are some degrees of subjective choice in the smoothing length. We have repeated the measurement varying the smoothing length over a reasonable range and found the difference of \numax\ values is less than $\sim 0.5\sigma$ (ranging from $1.28~\muHz$ to $1.34~\muHz$, with the value of \numax\ increasing as the smoothing length decreases). As for the two essential source properties, the source distance $D_{\rm s}$ and its angular radius $\theta_{\ast}$, the results are nearly the same within the uncertainties. Therefore, the choice of smoothing length does not have any significant impact on our results.

The power spectrum displays regular structure with a spacing of $\sim\,0.25\,\muHz$, consistent with the expected value for the large-frequency separation. However, due to aliasing and the fact that non-radial modes in high-luminosity giants have been shown to deviate from the asymptotic theory \citep{Stello2014ApJL}, we choose to only use \numax\ as a constraint in our analysis. We note that the validity of the \numax\ scaling relation for high-luminosity giants is still an active field of research. However, {\it Kepler} results have demonstrated that \numax\ remains a sensitive tracer of luminosity, connecting late K / early M giants to the the well-known period-luminosity relations in mid-to-late M giants \citep{Mosser2013A}.

The asteroseismic detection confirms that the microlensing source is an oscillating red giant. To characterize the source, we combined the seismic \numax\ measurement with the de-reddened NIR photometry $H_{\rm s,0} = 10.35 \pm 0.04$ and $K_{\rm s,0} = 10.11 \pm 0.04$ (cf.~Section~\ref{subsec:CMD}) to infer stellar parameters using \texttt{isoclassify} \citep{Huber2017ApJ}. The extinction value is determined from the CMD analysis (Section~\ref{subsec:CMD}), since the dust should almost all lie well in front of the bulge at $b=-1.84$. In summary, \texttt{isoclassify} uses a grid of MIST isochrones \citep{Choi2016ApJ} to probabilistically infer stellar parameters given any combination of photometric, spectroscopic or asteroseismic input parameters. The tight constraint on the evolutionary state from the \numax\ measurements enables us to constrain the radius and thus distance to the source to $\sim$\,20\,\% (Table~\ref{table:source}). From the distance estimate of $D_s\simeq 11.5$ kpc, one can infer that the source is most likely located beyond the bulge.

\subsection{CMD analysis}\label{subsec:CMD}
\begin{figure*}
\centering
\resizebox{\hsize}{!}{\includegraphics{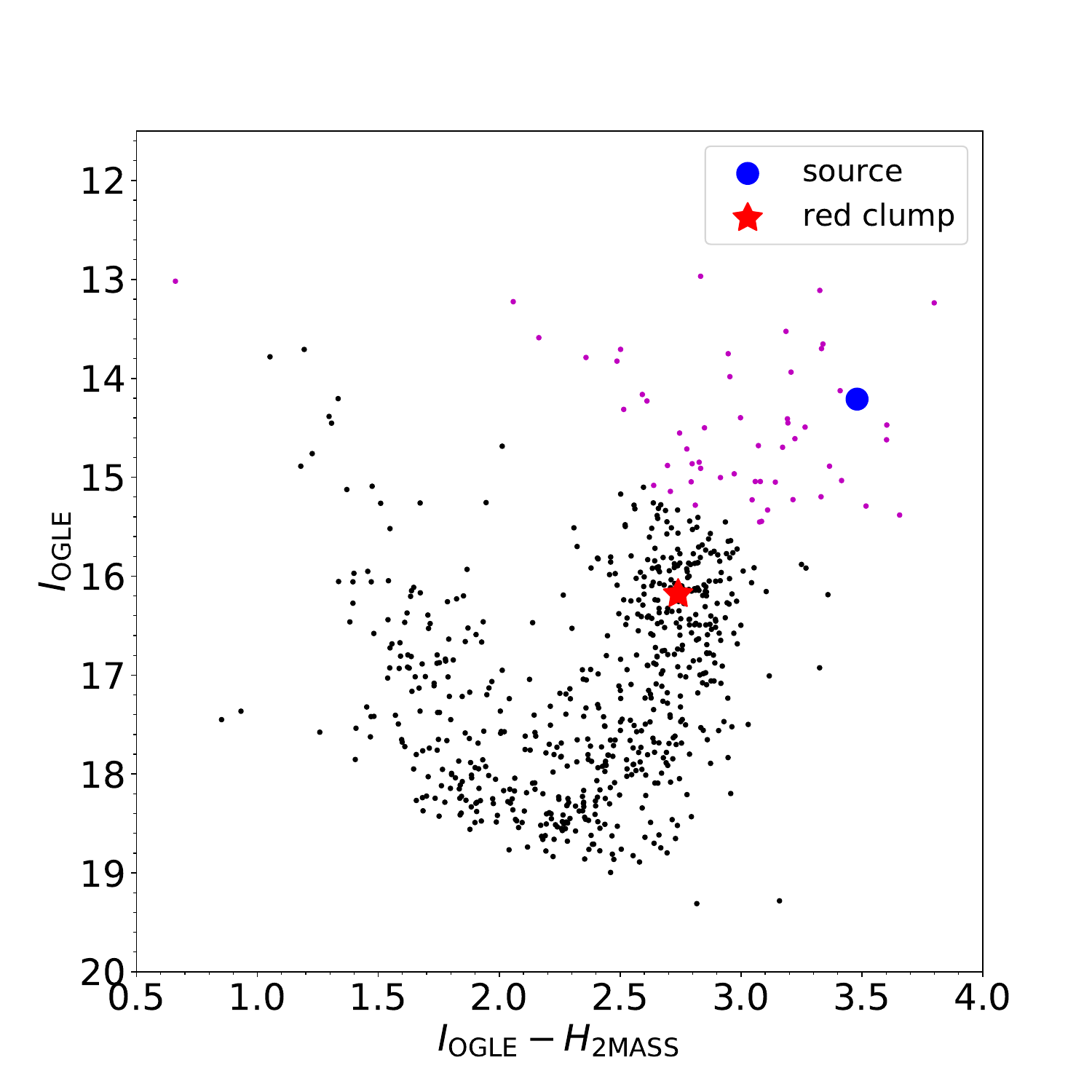}}
\caption{Colour-magnitude diagrams of a $2'\times 2'$ square centred around \target. Because the saturation limit of VVV catalogue is $H \sim 12.0$, we use $H$-band photometry from the VVV catalogue (black dots) for $H > 12.5$ and $H$-band photometry from the 2MASS catalogue (magenta dots) for $H < 12.5$. The VVV catalogue has been calibrated to the 2MASS photometric system. The red asterisk shows the centroid of the red clump, and the blue dot indicates the position of the source star.}
\label{fig:cmd}
\end{figure*}

We derive the extinction parameters by comparing the red-clump centroid on a CMD with its intrinsic brightness and colour. The $I - H$ versus $I$ CMD is constructed by cross-matching the OGLE-III~\citep{Udalski2008AcA} $I$-band stars with the VVV \citep{VVV} and the 2MASS $H$-band stars within a $2'\times 2'$ region centred around the event (See Figure \ref{fig:cmd}). The VVV catalogue is calibrated to the 2MASS photometric system.  We estimate the centroid of the red clump to be $(I - H, I)_{\rm cl} = (2.74 \pm 0.02, 16.18 \pm 0.03)$. By comparing it to the intrinsic value $(I - H, I)_{\rm cl,0} = (1.32, 14.36)$~\citep{Nataf2016}, we find the extinction and reddening to be $A_I = 1.82\pm0.03, E(I-H) = 1.42\pm0.03$. $A_I/E(I-H) = 1.28\pm0.04$, consistent with the extinction law of \cite{Nataf2016} and \cite{N09}. Using the extinction law $A_I : A_K = 7.26 : 1$  of \cite{Nataf2016}, we find that $A_K = 0.25\pm0.02$.

We can also estimate the angular radius $\theta_*$ of the source by placing the source on the CMD \citep{Albrow2000ApJ,Yoo2004ApJ}. The position of the source in the CMD is $(I - H, I)_{\rm s} = (3.49 \pm 0.03, 14.21 \pm 0.02)$ determined from the source OGLE $I$-band and CT13 $H$-band photometry. Assuming that the source suffers the same dust extinction as the red clump, its intrinsic position is $(I - H, I)_{\rm s,0} = (2.07 \pm 0.04, 12.39 \pm 0.04)$, which suggests that the source is an M2-giant star \citep{Bessell1998}. This is consistent with the effective temperature derived from the asteroseismic analysis. We convert the measured $I - H$ into $V - K$ using the colour-colour relation of \cite{Bessell1998} and then estimate the source angular radius using the CSB relation of M giants from \cite{Mgiant},
\begin{equation}\label{equ:cmd}
    \theta_* = 28.1 \pm 1.7~\mu {\rm as}~,
\end{equation}
which is consistent with the result obtained from the asteroseismic analysis~(Table~\ref{table:source}) within $\sim1 \sigma$. Due to its more precise constraint, we adopt the $\theta_{\ast}$ from the asteroseismology for further analyses.

\begin{table}
\begin{threeparttable}
\caption{Source properties}
\label{table:source}
\begin{tabularx}{\columnwidth}{l @{\extracolsep{\fill}} l}
\toprule
Parameter & Value \\
\midrule
 \vspace{1ex}
$I_{\rm s,0}$ & $12.39\pm0.04$ \\
 \vspace{1ex}
$(I - H)_{\rm s,0}$ & $2.07\pm0.04$ \\
 \vspace{1ex}
$(V - K)_{\rm s,0}$ & $4.39\pm0.07$ \\
\midrule
 \vspace{1ex}
Effective temperature, $T_{\mathrm{eff}}$ (K) & $3672^{+102}_{-93}$ \\
 \vspace{1ex}
Surface gravity, $\log~g$ (dex, cgs)  & $0.95^{+0.04}_{-0.05}$ \\
 \vspace{1ex}
Metallicity, $[Fe/H]$ & $0.01^{+0.15}_{-0.16}$ \\
 \vspace{1ex}
Mass, $M_{\mathrm{s}}$ ($M_{\odot}$) & $1.45^{+0.64}_{-0.38}$ \\
 \vspace{1ex}
Radius, $R_{\mathrm{s}}$ ($R_{\odot}$) & $67^{+13}_{-9}$ \\
 \vspace{1ex}
Distance, $D_{\mathrm{s}}$ (kpc) & $11.5^{+2.5}_{-1.7}$ \\
 \vspace{1ex}
Angular radius, $\theta_{\ast}$ ($\mu$as) & $26.9^{+0.50}_{-0.52}$ \\
\bottomrule
\end{tabularx}
    \begin{tablenotes}
      \small
      \item The magnitude and colour are estimated from the CMD analysis as described in Section~\ref{subsec:CMD}. Other values are derived from the asteroseismic analysis based on the OGLE-IV baseline data as described in Section~\ref{subsec:astero}.
    \end{tablenotes}
  \end{threeparttable}
\end{table}

\section{The lens: a Low-mass object beyond the foreground disc}\label{sec:lens}

Now that we have derived both the scaled source radius $\rho_{\ast}$ and the intrinsic source angular radius $\theta_{\ast}$, we can determine the Einstein radius using Equation~(\ref{equ:thetae}), $\theta_{\rm E} \simeq 0.094~{\rm mas}$. As already shown in Equation~(\ref{equ:m}), this, when combined with the parallax solutions from light curve modelling, can yield a lens mass measurement. In our case, there are two-degenerate mass solutions $M_{L}\simeq 0.073$ or $0.045~M_{\odot}$, making the lens either an ultracool dwarf or a brown dwarf. The distances to these two solutions are $D_L\simeq 9.8$ and $9.0$~kpc, respectively, where we have used Equation~(\ref{equ:d}), and the source distances are obtained from the asteroseismic analysis. The best-fit values with $1\sigma$ uncertainties are given in Table~\ref{table:lens}.

\subsection{Proper motion}\label{subsec:pm}
Owing to its high brightness, the source is in the \emph{Gaia} DR2 catalogue\footnote{The \emph{Gaia} source identifier is $4062797719417283456$.}~\citep{Gaia2016AA,Gaia2018AA}, with a proper-motion measurement: ${\bm \mu}_s(\alpha^{\ast},\delta)$\footnote{$\mu_{\alpha^{\ast}}(\equiv\mu_{\alpha}\cos\delta$) and $\mu_{\delta}$ are proper motions in right ascension and declination, respectively. The proper motion is measured in ICRS realized by an extragalactic
reference frame (\emph{Gaia}-CRF2), see \citet{Gaia2018AA2} for further details.} $=(-4.04,-7.49)\pm(0.27,0.20)$~mas~yr$^{-1}$, which equals $\bm{\mu}_s(l,b)=(-8.50,-0.23)\pm(0.22,0.25)$~mas~yr$^{-1}$.

Using the source distance obtained in Section~\ref{sec:source}, we can estimate the source velocity with respect to the Galactic centre as
\begin{equation}\label{equ:vs}
\bm{v}_s(l,b)=D_{s}\bm{\mu}_{s}+\bm{v}_{\sun}=(-222^{+100}_{-69},-6^{+14}_{-14})~{\rm km~s}^{-1}~,
\end{equation}
where $\bm{v}_{\sun}(l,b)=(241,7)$~km~s$^{-1}$ is the velocity of the Sun. We have adopted the disc rotation velocity $\bm{v}_{\rm rot}(l,b)=(229,0)$~km~s$^{-1}$ from \citet{Eilers2018arXiv} and the solar peculiar velocity $\bm{v}_{\sun,\rm pec}(l,b)=(12,7)$~km~s$^{-1}$ from \citet{Schonrich2010MNRAS}. By comparing the estimated source velocity with the disc rotation velocity, the conclusion that the source is a background disc star is further confirmed.

The lens velocity with respect to the Galactic centre can also be calculated through several procedures as detailed below.

First, the microlensing parameters derived from light-curve fitting directly give rise to the geocentric relative proper motion as
\begin{equation}\label{equ:pmrel}
\bm{\mu}_{\rm rel,geo} = \frac{\theta_{\rm E}}{t_{\rm E}}\frac{\bm{\pi}_E}{\pi_E}~.
\end{equation}
In order to meet the frame of $\bm{\mu}_s$, this should be transformed into the heliocentric frame by
\begin{equation}\label{equ:pmrelhel}
\bm{\mu}_{\rm rel,hel}=\bm{\mu}_{\rm rel,geo}+{\pi_{\rm rel}\over {\rm au}}\bm{v}_{\oplus,\perp}~,
\end{equation}
where $\bm{v}_{\oplus,\perp}(N,E)=(-1.07,25.43)$~km~s$^{-1}$ is Earth's projected velocity at the time of maximum magnification.

Now that we have both the source proper motion and the relative proper motion in the heliocentric frame, we can easily obtain the lens heliocentric proper motion by $\bm{\mu}_{L,\rm{hel}} = \bm{\mu}_s+\bm{\mu}_{\rm rel,hel}$. Lastly, the lens velocity with respect to the Galactic centre is derived by taking the Sun's motion into account as
\begin{equation}\label{equ:vl}
\bm{v}_{L} = D_{L}\bm{\mu}_{L,\rm{hel}}+\bm{v}_{\odot}~.
\end{equation}
Due to the four-fold degeneracy, there are four possible lens velocities, which are collected in Table~\ref{table:lens}.

\subsection{Location}
The estimated distances and proper motions all have somewhat large errors (see Table~\ref{table:lens}), resulting from the poorly constrained source distance (see Table~\ref{table:source}). Furthermore, two-degenerate distances are associated with four-degenerate proper motions.

There are three independent arguments that have the potential to break these degeneracies: (1) the $\chi^2$ values for each solution, (2) the `Rich argument'~\citep{Calchi2015ApJ}, and (3) a Bayesian analysis based on a Galactic model~\citep{Zhu2017AJ}. We here try to find the preferred combination of distance and proper motion by making use of each of these arguments.

For the $\chi^2$ values, as shown in Table~\ref{table:lens}, all solutions have comparable values, with differences $\lesssim5$. Considering potential systematic errors usually associated with light-curve modelling, these differences in $\chi^2$ are not large enough to rule out any solutions.

The `Rich argument', named after James Rich, is a statistical criterion based on the fact that, other things being equal, small parallax solutions are preferred over large ones by a factor of $(\pi_{\rm E, big}/\pi_{\rm E, small})^2$ ~\citep{Calchi2015ApJ,Calchi2018arXiv}. As the `Rich argument' is statistical in nature, it cannot be considered decisive for any given events especially when the difference between degenerate parallaxes is small~\citep{Ryu2018AJ,Calchi2018arXiv}. In our case, the `Rich argument' preference is only $\sim 2.6$, so it is also not large enough to favour any specific solutions.

The last argument is the Bayesian inference. Because the lens distance and velocity are calculated in each solution, we can statistically estimate the lens location for each solution by conducting a Bayesian analysis with a typical Galactic model~\citep{Zhu2017AJ,Wang2018ApJ}. In practice, we estimate the probabilities that the lens lies in the bulge or the disc separately in each solution. Combing the original bulge and disc probabilities of each solution with the Kroupa mass function~\citep{Kroupa2001MNRAS}, we can obtain the statistical weight (or the relative probability) of each solution. The results are shown in Table~\ref{table:lens}. Note that the reported bulge and disc probabilities are rescaled in each solution to make sure that the total probability is unity, and the relative probability is scaled according to the $(-,+)$ solution.

We also calculate the relative probabilities based on the $\chi^2$ values and the `Rich argument', respectively. Combing the three relative probabilities, we can obtain the total relative probability of each solution. As a result, there is no clear evidence for any preferred solution. In particular, the most interesting question is the relative probabilities of the $(+,+)$\&$(-,-)$ solutions vs.\ the $(+,-)$\&$(-,+)$ solutions. As can be seen in the last row of Table~\ref{table:lens}, these have almost equal probabilities ($0.359$ vs.\ $0.399$).  Combining the bulge and disc probabilities in each solution with its total relative probability, we conclude that the lens has a $\sim 35.1\%$ probability to be a bulge object and $\sim 64.9\%$ probability a background disc object.

\begin{table*}
\begin{threeparttable}
\caption{Lens properties}
\label{table:lens}
\begin{tabular}{lrrrrrrrr}
\toprule
 &
\multicolumn{1}{c}{$(+,+)$} &
\multicolumn{1}{c}{$(+,-)$} &
\multicolumn{1}{c}{$(-,+)$} &
\multicolumn{1}{c}{$(-,-)$} \\
\midrule
\vspace{1ex}
$\chi^2 / N_{\rm data}$	&
11561.5 / 11577	&
11556.3 / 11577	&
11561.3 / 11577	&
11558.5 / 11577	\\
 \vspace{1ex}
$\pi_{\rm E}$	&
$0.1581^{+0.0019}_{-0.0017}$	&
$0.2551^{+0.0026}_{-0.0027}$	&
$0.2542^{+0.0024}_{-0.0023}$	&
$0.1593^{+0.0019}_{-0.0021}$	\\
 \vspace{1ex}
$\theta_{\rm E}$ (mas)	&
$0.0939^{+0.0018}_{-0.0019}$       &
$0.0939^{+0.0018}_{-0.0019}$       &
$0.0939^{+0.0018}_{-0.0019}$       &
$0.0940^{+0.0018}_{-0.0019}$       \\
 \vspace{1ex}
$M_{L}$ ($M_{\odot}$)          &
$0.0730^{+0.0016}_{-0.0017}$       &
$0.0452^{+0.0010}_{-0.0010}$       &
$0.0454^{+0.0010}_{-0.0010}$       &
$0.0725^{+0.0016}_{-0.0017}$       \\
 \vspace{1ex}
$D_{L}$ (kpc)                     &
$9.8^{+1.8}_{-1.2}$           &
$9.0^{+1.5}_{-1.0}$           &
$9.0^{+1.5}_{-1.0}$           &
$9.8^{+1.8}_{-1.2}$          \\
 \vspace{1ex}
$\mu_{\rm{rel},\rm{geo},N}$ (mas yr$^{-1}$)     &
$1.960^{+0.058}_{-0.056}$           &
$-2.429^{+0.059}_{-0.062}$           &
$2.351^{+0.057}_{-0.057}$           &
$-2.087^{+0.060}_{-0.065}$          \\
 \vspace{1ex}
$\mu_{\rm{rel},\rm{geo},E}$ (mas yr$^{-1}$)     &
$-1.728^{+0.043}_{-0.044}$           &
$-0.964^{+0.025}_{-0.025}$           &
$-1.143^{+0.027}_{-0.028}$           &
$-1.573^{+0.041}_{-0.043}$          \\
 \vspace{1ex}
$\mu_{L,\rm{hel},N}$ (mas yr$^{-1}$)     &
$-5.53^{+0.21}_{-0.21}$           &
$-9.92^{+0.21}_{-0.21}$           &
$-5.14^{+0.21}_{-0.21}$           &
$-9.58^{+0.21}_{-0.21}$          \\
 \vspace{1ex}
$\mu_{L,\rm{hel},E}$ (mas yr$^{-1}$)     &
$-5.68^{+0.27}_{-0.27}$           &
$-4.87^{+0.27}_{-0.27}$          &
$-5.05^{+0.27}_{-0.27}$          &
$-5.53^{+0.27}_{-0.27}$         \\
 \vspace{1ex}
$v_{L,l}$ (km s$^{-1}$)     &
$-114^{+65}_{-46}$           &
$-230^{+80}_{-55}$           &
$-57^{+51}_{-36}$           &
$-273^{+94}_{-66}$          \\
 \vspace{1ex}
$v_{L,b}$ (km s$^{-1}$)     &
$108^{+22}_{-17}$           &
$-24^{+12}_{-11}$           &
$85^{+17}_{-14}$           &
$8^{+12}_{-12}$          \\
\midrule
Bulge Prob$^a$     &
$0.963$           &
$0.343$           &
$0.997$           &
$0.238$          \\
disc Prob$^a$     &
$0.037$           &
$0.657$           &
$0.003$           &
$0.762$          \\
\midrule
Relative Prob (Bayesian inference)     &
$0.424$           &
$0.957$           &
$1.000$           &
$0.999$          \\
Relative Prob ($\chi^2$)     &
$0.074$           &
$1.000$           &
$0.082$           &
$0.333$          \\
Relative Prob (Rich argument)   &
$1.000$           &
$0.384$           &
$0.387$           &
$0.985$          \\
\midrule
Relative Prob (total)    &
$0.031$           &
$0.367$           &
$0.032$           &
$0.328$          \\
\bottomrule
\end{tabular}
    \begin{tablenotes}
      \small
      \item All the results are determined based on the microlensing parameters obtained from Traditional method.
      \item $^a$ Location probability for each solution is scaled to make the total probability unity.
    \end{tablenotes}
  \end{threeparttable}
\end{table*}

\section{Summary and discussion}\label{sec:dis}

We present the analysis of the microlensing event~\target, which has both the finite-source effects and the space-based microlens parallax detected. There are two degenerate solutions to the event. In one case, the lens is a brown dwarf with a mass $\simeq 0.045~M_{\odot}$ located at $D_L\simeq 9.0$~kpc. In the other case, it is an ultracool dwarf with $M_L\simeq 0.073~M_{\odot}$ located at $D_L\simeq 9.8$~kpc. We have tried to break the degeneracy by adopting three independent arguments: the difference in $\chi^2$ between the four solutions, the `Rich argument', and Bayesian inference. None of these arguments is strong enough to choose a specific solution. Specifically, the solutions are essentially degenerate, with $\delta \chi^2 \lesssim 5$. The "Rich argument"  only favours the best solution by a factor of $\sim2.6$. With a typical Galactic model and the Kroupa mass function, the Bayesian analysis disfavours the $(+,+)$ solution by a factor of $\gtrsim 2.4$. Combining these probabilities, we find that the solution that places the lens is in the background disc is only slightly favoured, with a $\sim64.9\%$ probability. The geocentric relative proper motion of this system is around $2.6~{\rm mas~yr^{-1}}$, so the separation between the lens and the source will be around $26$ mas in 2027, which can be resolved by the next generation telescopes ($D \sim 30$~m class telescopes, such as E-ELT, TMT and GMT, have a resolution $\theta \sim 14(D/30~{\rm m})^{-1}$ mas in $H$-band). In other words, the degeneracy can be broken at first light of 30~m telescopes.

The source star is a bright oscillating red giant, making it possible to use OGLE baseline data to perform the asteroseismic analysis. Asteroseismic detection reveals that the source is located in the background disc with $D_s\simeq11.5^{+2.5}_{-1.7}$~kpc. To our knowledge, this is the first microlensing system whose source is unambiguously identified as a background disc star.\footnote{Although the system reported in \citet{Shvartzvald2018ApJ} is also expected to have a source residing in the background disc, its distance estimation is quite uncertain due to the complex extinction.} The measurement of the source distance enables us to determine the lens distance. Customarily, microlens source stars are assumed to reside in the Galactic bulge, due to the higher lensing rate of bulge stars compared to disc stars. While this argument is acceptable for the majority of high-latitude events ($|b|\gtrsim2$), it does not necessarily hold when dealing with events located at low latitude, where the disc population can contribute more to the microlens sources. Actually, several far disc sources have already been claimed in the recent published low-latitude events~\citep{Shvartzvald2018ApJ,Bennett2018AJ}. Furthermore, low-latitude events suffer more complex extinction due to spatial and radial non-uniform reddening. Both complex extinction and uncertainty of the source distance can affect the accuracy of estimated lens physical properties.

These issues will become a concern for the proposed \emph{WFIRST} microlensing survey, because the target fields are at low latitude to take advantage of the higher event rate~\citep{Gould1995ApJ2,Shvartzvald2017AJ,Navarro2018ApJ,Penny2018arXiv}. The asteroseismic analysis conducted in this paper provides an opportunity to ease this tension. The ability and accuracy of asteroseismology in determining star properties and distances have been well tested in previous statistical studies (see, e.g., \citealt{Huber2017ApJ}). In our practice, although only one asteroseismic parameter $\nu_{\rm max}$ is measured, the constraint on the stellar evolutionary state enables us to constrain the source distance to $\sim 20\%$. This can be improved if that continued monitoring of the source is achieved (see, e.g., \citealt{Hekker2012A}).

Another important improvement brought by the asteroseismology is in the aspect of the source angular size measurement. Even in this case in which only one asteroseismic parameter $\nu_{\rm max}$ is determined, the constraint on the angular radius is already $\sim 3$ times better than that derived from empirical colour-surface brightness relations (see Table~\ref{table:source} and Equation~\ref{equ:cmd}). When a better photometric precision is obtained in the near future using observations with \emph{WFIRST}, a complete asteroseismic measurement is expected to provide better constraints on the angular sizes of the sources of microlensing events. \citep{Gould2015JKAS}.

A comparison of the angular radius derived from asteroseismic analysis and from traditional CMD analysis also enables us to test the empirical CSB relation of M giants. Usually, the CSB relation from \cite{Kervella2004b} is used in microlensing analyses to estimate the source angular radius \citep[e.g.,][]{MB15020,MrozNeptune,Mroz2FFP}. Using this method, we derived an angular radius of $32.4 \pm 1.9~\mu {\rm as}$, $\sim2\sigma$ away from that from asteroseismic analysis. This discrepancy is not unexpected, given that the CSB relation in \cite{Kervella2004b} was derived from giants with colour $(V - K)_{0} < 2.5$, while the source here has a much redder colour $(V - K)_0 \sim 4.40$. Actually, \cite{Mgiant} indicated that the CSB relation of M giants is different from that of other-type giants. If we adopt the new CSB relation of M giants from \cite{Mgiant}, the result is consistent with that derived from asteroseismic analysis within $1\sigma$ (see Section~\ref{sec:source}). In this aspect, our results lend some support to their claims.

Although few, M giants play important roles in single events, as finite-source effects are strongly biased toward large (hence, bright and red) source stars \citep{Shvartzvald2018arXiv}. In fact, the three free-floating planet candidates with finite-source effects measured all have M-giant sources \citep{MrozNeptune, Mroz2FFP}. \cite{Mroz2FFP} finds that the CSB relation of M giants from \cite{Mgiant} gives angular radii that are systematically $10\%$ lower than those derived from \cite{Kervella2004b}. Hence, this should be noted when estimating the angular radii of M-giant sources.

A side effect of variable stars in microlensing surveys is that they always incur correlated noise, which can affect the accuracy of light-curve modelling or even mimic microlensing events. Hence, a selection criterion of constant baseline is often applied in microlensing search algorithms. This is obviously not an optimal strategy in consideration of the ubiquity and the value of variable stars. A possible solution is to introduce the Gaussian Processes (GPs) to handle the correlated noise. In the literature of transit timing analysis and radial velocity measurements, light-curve analyses have already been well equipped with the GP method (see, e.g., \citealt{Gibson2012MNRAS,Brewer2009MNRAS,Grunblatt2016AJ,Czekala2017ApJ}). In this paper, we have tested this technique in modelling the event~\target. The feasibility of GP method in tackling the correlated noise is noticeable from the reduction of the size of the residuals from the model fit to the light curve (Figure~\ref{fig:lc}), even though we only used the simplest strategy. Nevertheless, as already mentioned in Section~\ref{subsec:GP}, there are still some unsolved problems in the practical aspects of the GP model for the microlensing analysis. Among them the most urgent ones are exploring strategies to tackle blending effects and error rescaling. An exploration of how the GP model hyperparameters and the microlensing parameters are correlated is also necessary. All these issues can be treated by careful numerical experiments using mock microlensing events with different microlensing parameters. We defer such a study to future work.

\section*{Acknowledgements}

S.-S.L., W.Z. and S.M. acknowledge support by the National Science Foundation of China (Grant No.~11821303 and 11761131004). The OGLE has received funding from the National Science Centre, Poland, grant MAESTRO 2014/14/A/ST9/00121 to A.U.. Work by Y.S. was supported by an appointment to the NASA Postdoctoral Program at the Jet Propulsion Laboratory, California Institute of Technology, administered by Universities Space Research Association through a contract with NASA. D.H. acknowledges support by the National Science Foundation (AST-1717000). This research has made use of the KMTNet system operated by the Korea Astronomy and Space Science Institute (KASI) and the data were obtained at three host sites of CTIO in Chile, SAAO in South Africa, and SSO in Australia. The MOA project is supported by JSPS KAKENHI Grant Number JSPS24253004, JSPS26247023, JSPS23340064, JSPS15H00781, JP16H06287, and JP17H02871. Work by A.G. was supported by AST-1516842 from the US NSF and JPL grant 1500811. Work by P.F. and W.Z. was supported by Canada-France-Hawaii Telescope (CFHT). S.D. acknowledges Projects 11573003 supported by the National Science Foundation of China (NSFC). Work by C.H. was supported by the grant (2017R1A4A1015178) of National Research Foundation of Korea. The research has made use of data obtained at the Danish 1.54~m telescope at ESOs La Silla Observatory. Y.T. acknowledges the support of DFG priority program SPP 1992 `Exploring the Diversity of Extrasolar Planets' (WA 1047/11-1). CITEUC is funded by National Funds through FCT - Foundation for Science and Technology (project: UID/Multi/00611/2013) and FEDER - European Regional Development Fund through COMPETE 2020 - Operational Programme Competitiveness and Internationalization (project: POCI-01-0145-FEDER-006922). L.M. acknowledges support from the Italian Minister of Instruction, University and Research (MIUR) through FFABR 2017 fund. J.-P.B. was supported by the University of Tasmania through the UTAS Foundation and the endowed Warren Chair in Astronomy. J.-P.B. acknowledges the financial support of the CNES and the ANR COLD-WORLDS, ANR-18-CE31-0002. Development of the Greenhill Observatory was supported under the Australian Research Council's LIEF funding scheme (project LE110100055). We thank C. Harlingten for the use of the H127 Telescope.

This work has made use of data from the European Space Agency (ESA) mission
{\it Gaia} (\url{https://www.cosmos.esa.int/gaia}), processed by the {\it Gaia}
Data Processing and Analysis Consortium (DPAC,
\url{https://www.cosmos.esa.int/web/gaia/dpac/consortium}). Funding for the DPAC
has been provided by national institutions, in particular the institutions
participating in the {\it Gaia} Multilateral Agreement.

This publication makes use of data products from the Two Micron All Sky Survey, which is a joint project of the University of Massachusetts and the Infrared Processing and Analysis Center/California Institute of Technology, funded by the National Aeronautics and Space Administration and the National Science Foundation.


\vspace{0.25cm}

\noindent 
$^{1}$National Astronomical Observatories, Chinese Academy of Sciences, Beijing 100101, China\\
$^{2}$School of Astronomy and Space Science, University of Chinese Academy of Sciences, Beijing 100049, China\\
$^{3}$Physics Department and Tsinghua Centre for Astrophysics, Tsinghua University, Beijing 100084, China\\
$^{4}$Warsaw University Observatory, Al.~Ujazdowskie~4, 00-478~Warszawa, Poland\\
$^{5}$IPAC, Mail Code 100-22, Caltech, 1200 E. California Blvd., Pasadena, CA 91125, USA\\
$^{6}$Institute for Astronomy, University of Hawaii, 2680 Woodlawn Drive, Honolulu, HI 96822, USA\\
$^{7}$Korea Astronomy and Space Science Institute, Daejon 34055, Republic of Korea\\
$^{8}$University of Science and Technology, Korea, (UST), 217 Gajeong-ro Yuseong-gu, Daejeon 34113, Korea\\
$^{9}$Department of Earth and Space Science, Graduate School of Science, Osaka University, Toyonaka, Osaka 560-0043, Japan\\
$^{10}$Department of Astronomy, Ohio State University, 140 W. 18th Ave., Columbus, OH~43210,~USA\\
$^{11}$Max Planck Institute for Astronomy, K{\"o}nigstuhl 17, 69117 Heidelberg, Germany\\
$^{12}$CFHT Corporation, 65-1238 Mamalahoa Hwy, Kamuela, Hawaii 96743, USA\\
$^{13}$Universit\'e de Toulouse, UPS-OMP, IRAP, Toulouse, France\\
$^{14}$Kavli Institute for Astronomy and Astrophysics, Peking University, Yi He Yuan Road 5, Hai Dian District, Beijing 100871, China\\
$^{15}$Niels Bohr Institute \& Centre for Star and Planet Formation, University of Copenhagen, {\O}ster Voldgade 5, 1350 Copenhagen, Denmark\\
$^{16}$School of Natural Sciences, University of Tasmania, Private Bag 37 Hobart, Tasmania 7001 Australia\\
$^{17}$Department of Physics, University of Warwick, Gibbet Hill Road, Coventry, CV4~7AL,~UK\\
$^{18}$Center for Astrophysics | Harvard \& Smithsonian, 60 Garden St.,Cambridge, MA 02138, USA\\
$^{19}$Jet Propulsion Laboratory, California Institute of Technology, 4800 Oak Grove Drive, Pasadena, CA 91109, USA\\
$^{20}$Canadian Institute for Theoretical Astrophysics, University of Toronto, 60 St George Street, Toronto, ON M5S 3H8, Canada\\
$^{21}$University of Canterbury, Department of Physics and Astronomy, Private Bag 4800, Christchurch 8020, New Zealand\\
$^{22}$Department of Physics, Chungbuk National University, Cheongju 28644, Republic of Korea\\
$^{23}$School of Space Research, Kyung Hee University, Yongin, Kyeonggi 17104, Republic of Korea\\
$^{24}$Center for Cosmology \& AstroParticle Physics, The Ohio State University, 191 West Woodruff Avenue, Columbus, OH~43210,~USA\\
$^{25}$Institute of Natural and Mathematical Sciences, Massey University, Auckland 0745, New Zealand\\
$^{26}$Institute for Space-Earth Environmental Research, Nagoya University, Nagoya 464-8601, Japan\\
$^{27}$Code 667, NASA Goddard Space Flight Center, Greenbelt, MD 20771, USA\\
$^{28}$Department of Astronomy, University of Maryland, College Park, MD 20742, USA\\
$^{29}$Department of Physics, University of Auckland, Private Bag 92019, Auckland, New Zealand\\
$^{30}$Okayama Astrophysical Observatory, National Astronomical Observatory of Japan, 3037-5 Honjo, Kamogata, Asakuchi, Okayama 719-0232, Japan\\
$^{31}$Department of Astronomy, Graduate School of Science, The University of Tokyo, 7-3-1 Hongo, Bunkyo-ku, Tokyo 113-0033, Japan\\
$^{32}$National Astronomical Observatory of Japan, 2-21-1 Osawa, Mitaka, Tokyo 181-8588, Japan\\
$^{33}$School of Chemical and Physical Sciences, Victoria University, Wellington, New Zealand\\
$^{34}$Institute of Space and Astronautical Science, Japan Aerospace Exploration Agency, 3-1-1 Yoshinodai, Chuo, Sagamihara, Kanagawa, 252-5210, Japan\\
$^{35}$University of Canterbury Mt.\ John Observatory, P.O. Box 56, Lake Tekapo 8770, New Zealand\\
$^{36}$Department of Physics, Faculty of Science, Kyoto Sangyo University, 603-8555 Kyoto, Japan\\
$^{37}$Auckland Observatory, Auckland, New Zealand\\
$^{38}$Possum Observatory, Patutahi, New Zealand\\
$^{39}$Kumeu Observatory, Kumeu, New Zealand\\
$^{40}$Institute for Radio Astronomy and Space Research (IRASR), AUT University, Auckland, New Zealand\\
$^{41}$Turitea Observatory, Palmerston North, New Zealand\\
$^{42}$Las Cumbres Observatory, 6740 Cortona Drive, suite 102, Goleta, CA 93117, USA\\
$^{43}$School of Physics and Astronomy, Tel-Aviv University, Tel-Aviv 6997801, Israel\\
$^{44}$Astronomisches Rechen-Institut, Zentrum f{\"u}r Astronomie der Universit{\"a}t Heidelberg (ZAH), 69120 Heidelberg, Germany\\
$^{45}$Dipartimento di Fisica ``E.R. Caianiello'', Universit{\`a} di Salerno, Via Giovanni Paolo II 132, 84084, Fisciano, Italy\\
$^{46}$Istituto Nazionale di Fisica Nucleare, Sezione di Napoli, Napoli, Italy\\
$^{47}$Centre for Exoplanet Science, SUPA, School of Physics \& Astronomy, University of St Andrews, North Haugh, St Andrews KY16 9SS, UK\\
$^{48}$CITEUC -- Center for Earth and Space Research of the University of Coimbra, Geophysical and Astronomical Observatory, R. Observat{\'o}rio s/n, 3040-004 Coimbra, Portugal\\
$^{49}$Department of Physics, Isfahan University of Technology, Isfahan 84156-83111, Iran\\
$^{50}$Universit{\"a}t Hamburg, Faculty of Mathematics, Informatics and Natural Sciences, Department of Earth Sciences, Meteorological Institute, Bundesstra\ss{}e 55, 20146 Hamburg, Germany\\
$^{51}$Astrophysics Group, Keele University, Staffordshire, ST5 5BG, UK\\
$^{52}$Institute for Advanced Research and Dept.\, of Physics, Nagoya University, Furo-cho, Chikusa-ku, Nagoya, 464-8601, Japan\\
$^{53}$Instituto de Astronomia y Ciencias Planetarias de Atacama,  Universidad de Atacama, Copayapu 485,  Copiapo, Chile\\
$^{54}$Chungnam National University, Department of Astronomy and Space Science, 34134 Daejeon, Republic of Korea\\
$^{55}$Dipartimento di Fisica, Universit{\`a} di Roma Tor Vergata, Via della Ricerca Scientifica 1, I-00133-Roma, Italy\\
$^{56}$International Institute for Advanced Scientific Studies (IIASS), Via G. Pellegrino 19, I-84019, Vietri sul Mare (SA), Italy\\
$^{57}$Centro de Astronom{\'{i}}a, Universidad de Antofagasta, Av.\ Angamos 601, Antofagasta, Chile\\
$^{58}$Department of Physics, Sharif University of Technology, PO Box 11155-9161 Tehran, Iran\\
$^{59}$Las Cumbres Observatory Global Telescope, 6740 Cortona Dr., Suite 102, Goleta, CA 93111, USA\\
$^{60}$Department of Physics, University of California, Santa Barbara, CA 93106-9530, USA\\
$^{61}$Centre for Electronic Imaging, Department of Physical Sciences, The Open University, Milton Keynes, MK7 6AA, UK\\
$^{62}$Institute for Astronomy, University of Edinburgh, Royal Observatory, Edinburgh EH9 3HJ, UK\\
$^{63}$Stellar Astrophysics Centre, Department of Physics and Astronomy, Aarhus University, Ny Munkegade 120, 8000 Aarhus C, Denmark\\
$^{64}$Sorbonne Universit\'es, UPMC Universit\'e Paris 6 et CNRS, UMR 7095, Institut dAstrophysique de Paris, 98 bis bd Arago, F-75014 Paris, France\\
$^{A}$The OGLE Collaboration\\
$^{B}$The \emph{Spitzer} Team\\
$^{C}$The KMTNet Collaboration\\
$^{D}$The MOA Collaboration\\
$^{E}$The LCO and $\mu$FUN Follow-up Teams\\
$^{F}$The MiNDSTEp Collaboration\\
$^{G}$The Tasmanian Telescope



\bibliographystyle{mnras}
\bibliography{reference}

\begin{thebibliography}{}
\makeatletter
\relax
\def\mn@urlcharsother{\let\do\@makeother \do\$\do\&\do\#\do\^\do\_\do\%\do\~}
\def\mn@doi{\begingroup\mn@urlcharsother \@ifnextchar [ {\mn@doi@}
  {\mn@doi@[]}}
\def\mn@doi@[#1]#2{\def\@tempa{#1}\ifx\@tempa\@empty \href
  {http://dx.doi.org/#2} {doi:#2}\else \href {http://dx.doi.org/#2} {#1}\fi
  \endgroup}
\def\mn@eprint#1#2{\mn@eprint@#1:#2::\@nil}
\def\mn@eprint@arXiv#1{\href {http://arxiv.org/abs/#1} {{\tt arXiv:#1}}}
\def\mn@eprint@dblp#1{\href {http://dblp.uni-trier.de/rec/bibtex/#1.xml}
  {dblp:#1}}
\def\mn@eprint@#1:#2:#3:#4\@nil{\def\@tempa {#1}\def\@tempb {#2}\def\@tempc
  {#3}\ifx \@tempc \@empty \let \@tempc \@tempb \let \@tempb \@tempa \fi \ifx
  \@tempb \@empty \def\@tempb {arXiv}\fi \@ifundefined
  {mn@eprint@\@tempb}{\@tempb:\@tempc}{\expandafter \expandafter \csname
  mn@eprint@\@tempb\endcsname \expandafter{\@tempc}}}

\bibitem[\protect\citeauthoryear{{Aerts}, {Christensen-Dalsgaard}  \&
  {Kurtz}}{{Aerts} et~al.}{2010}]{Aerts2010aste.book}
{Aerts} C.,  {Christensen-Dalsgaard} J.,   {Kurtz} D.~W.,  2010,
  {Asteroseismology}

\bibitem[\protect\citeauthoryear{{Alard} \& {Lupton}}{{Alard} \&
  {Lupton}}{1998}]{Alard1998}
{Alard} C.,  {Lupton} R.~H.,  1998, \mn@doi [\apj] {10.1086/305984}, \href
  {http://adsabs.harvard.edu/abs/1998ApJ...503..325A} {503, 325}

\bibitem[\protect\citeauthoryear{{Albrow} et~al.,}{{Albrow}
  et~al.}{2000}]{Albrow2000ApJ}
{Albrow} M.~D.,  et~al., 2000, \mn@doi [\apj] {10.1086/308798}, \href
  {http://adsabs.harvard.edu/abs/2000ApJ...534..894A} {534, 894}

\bibitem[\protect\citeauthoryear{{Albrow} et~al.,}{{Albrow}
  et~al.}{2009}]{pysis}
{Albrow} M.~D.,  et~al., 2009, \mn@doi [\mnras]
  {10.1111/j.1365-2966.2009.15098.x}, \href
  {http://adsabs.harvard.edu/abs/2009MNRAS.397.2099A} {397, 2099}

\bibitem[\protect\citeauthoryear{{An} et~al.,}{{An} et~al.}{2002}]{An2002ApJ}
{An} J.~H.,  et~al., 2002, \mn@doi [\apj] {10.1086/340191}, \href
  {http://adsabs.harvard.edu/abs/2002ApJ...572..521A} {572, 521}

\bibitem[\protect\citeauthoryear{{Barclay}, {Endl}, {Huber}, {Foreman-Mackey},
  {Cochran}, {MacQueen}, {Rowe}  \& {Quintana}}{{Barclay}
  et~al.}{2015}]{Barclay2015ApJ}
{Barclay} T.,  {Endl} M.,  {Huber} D.,  {Foreman-Mackey} D.,  {Cochran} W.~D.,
  {MacQueen} P.~J.,  {Rowe} J.~F.,   {Quintana} E.~V.,  2015, \mn@doi [\apj]
  {10.1088/0004-637X/800/1/46}, \href
  {http://adsabs.harvard.edu/abs/2015ApJ...800...46B} {800, 46}

\bibitem[\protect\citeauthoryear{{Bennett} et~al.,}{{Bennett}
  et~al.}{2012}]{Bennett2012ApJ}
{Bennett} D.~P.,  et~al., 2012, \mn@doi [\apj] {10.1088/0004-637X/757/2/119},
  \href {http://adsabs.harvard.edu/abs/2012ApJ...757..119B} {757, 119}

\bibitem[\protect\citeauthoryear{{Bennett} et~al.,}{{Bennett}
  et~al.}{2018}]{Bennett2018AJ}
{Bennett} D.~P.,  et~al., 2018, \mn@doi [\aj] {10.3847/1538-3881/aad59c}, \href
  {http://adsabs.harvard.edu/abs/2018AJ....156..113B} {156, 113}

\bibitem[\protect\citeauthoryear{{Bessell} \& {Brett}}{{Bessell} \&
  {Brett}}{1988}]{Bessell1998}
{Bessell} M.~S.,  {Brett} J.~M.,  1988, \mn@doi [\pasp] {10.1086/132281}, \href
  {http://adsabs.harvard.edu/abs/1988PASP..100.1134B} {100, 1134}

\bibitem[\protect\citeauthoryear{{Bond} et~al.,}{{Bond}
  et~al.}{2001}]{Bond2001}
{Bond} I.~A.,  et~al., 2001, \mn@doi [\mnras]
  {10.1046/j.1365-8711.2001.04776.x}, \href
  {http://adsabs.harvard.edu/abs/2001MNRAS.327..868B} {327, 868}

\bibitem[\protect\citeauthoryear{{Boyajian}, {van Belle}  \& {von
  Braun}}{{Boyajian} et~al.}{2014}]{Boyajian2014AJ}
{Boyajian} T.~S.,  {van Belle} G.,   {von Braun} K.,  2014, \mn@doi [\aj]
  {10.1088/0004-6256/147/3/47}, \href
  {http://adsabs.harvard.edu/abs/2014AJ....147...47B} {147, 47}

\bibitem[\protect\citeauthoryear{{Bramich}}{{Bramich}}{2008}]{DanDIA}
{Bramich} D.~M.,  2008, \mn@doi [\mnras] {10.1111/j.1745-3933.2008.00464.x},
  \href {http://adsabs.harvard.edu/abs/2008MNRAS.386L..77B} {386, L77}

\bibitem[\protect\citeauthoryear{{Brewer} \& {Stello}}{{Brewer} \&
  {Stello}}{2009}]{Brewer2009MNRAS}
{Brewer} B.~J.,  {Stello} D.,  2009, \mn@doi [\mnras]
  {10.1111/j.1365-2966.2009.14679.x}, \href
  {http://adsabs.harvard.edu/abs/2009MNRAS.395.2226B} {395, 2226}

\bibitem[\protect\citeauthoryear{{Brown} \& {Gilliland}}{{Brown} \&
  {Gilliland}}{1994}]{Brown1994ARAA}
{Brown} T.~M.,  {Gilliland} R.~L.,  1994, \mn@doi [\araa]
  {10.1146/annurev.aa.32.090194.000345}, \href
  {http://adsabs.harvard.edu/abs/1994ARA%26A..32...37B} {32, 37}

\bibitem[\protect\citeauthoryear{{Brown}, {Gilliland}, {Noyes}  \&
  {Ramsey}}{{Brown} et~al.}{1991}]{Brown1991ApJ}
{Brown} T.~M.,  {Gilliland} R.~L.,  {Noyes} R.~W.,   {Ramsey} L.~W.,  1991,
  \mn@doi [\apj] {10.1086/169725}, \href
  {http://adsabs.harvard.edu/abs/1991ApJ...368..599B} {368, 599}

\bibitem[\protect\citeauthoryear{{Calchi Novati} et~al.,}{{Calchi Novati}
  et~al.}{2015a}]{Calchi2015ApJ}
{Calchi Novati} S.,  et~al., 2015a, \mn@doi [\apj]
  {10.1088/0004-637X/804/1/20}, \href
  {http://adsabs.harvard.edu/abs/2015ApJ...804...20C} {804, 20}

\bibitem[\protect\citeauthoryear{{Calchi Novati} et~al.,}{{Calchi Novati}
  et~al.}{2015b}]{Spitzerdata}
{Calchi Novati} S.,  et~al., 2015b, \mn@doi [\apj]
  {10.1088/0004-637X/814/2/92}, \href
  {http://adsabs.harvard.edu/abs/2015ApJ...814...92C} {814, 92}

\bibitem[\protect\citeauthoryear{{Calchi Novati} et~al.,}{{Calchi Novati}
  et~al.}{2018}]{Calchi2018arXiv}
{Calchi Novati} S.,  et~al., 2018, arXiv e-prints, \href
  {https://ui.adsabs.harvard.edu/\#abs/2018arXiv180105806C} {p.
  arXiv:1801.05806}

\bibitem[\protect\citeauthoryear{{Carpenter}}{{Carpenter}}{2001}]{2MASS}
{Carpenter} J.~M.,  2001, \mn@doi [\aj] {10.1086/320383}, \href
  {http://adsabs.harvard.edu/abs/2001AJ....121.2851C} {121, 2851}

\bibitem[\protect\citeauthoryear{{Carter} et~al.,}{{Carter}
  et~al.}{2012}]{Carter2012Sci}
{Carter} J.~A.,  et~al., 2012, \mn@doi [Science] {10.1126/science.1223269},
  \href {http://adsabs.harvard.edu/abs/2012Sci...337..556C} {337, 556}

\bibitem[\protect\citeauthoryear{{Chaplin} et~al.,}{{Chaplin}
  et~al.}{2014}]{Chaplin2014ApJS}
{Chaplin} W.~J.,  et~al., 2014, \mn@doi [\apjs] {10.1088/0067-0049/210/1/1},
  \href {http://adsabs.harvard.edu/abs/2014ApJS..210....1C} {210, 1}

\bibitem[\protect\citeauthoryear{{Choi} et~al.,}{{Choi}
  et~al.}{2012}]{Choi2012ApJ}
{Choi} J.-Y.,  et~al., 2012, \mn@doi [\apj] {10.1088/0004-637X/751/1/41}, \href
  {http://adsabs.harvard.edu/abs/2012ApJ...751...41C} {751, 41}

\bibitem[\protect\citeauthoryear{{Choi}, {Dotter}, {Conroy}, {Cantiello},
  {Paxton}  \& {Johnson}}{{Choi} et~al.}{2016}]{Choi2016ApJ}
{Choi} J.,  {Dotter} A.,  {Conroy} C.,  {Cantiello} M.,  {Paxton} B.,
  {Johnson} B.~D.,  2016, \mn@doi [\apj] {10.3847/0004-637X/823/2/102}, \href
  {http://adsabs.harvard.edu/abs/2016ApJ...823..102C} {823, 102}

\bibitem[\protect\citeauthoryear{{Christensen-Dalsgaard}}{{Christensen-Dalsgaard}}{2004}]{Christensen2004SoPh}
{Christensen-Dalsgaard} J.,  2004, \mn@doi [\solphys]
  {10.1023/B:SOLA.0000031392.43227.7d}, \href
  {http://adsabs.harvard.edu/abs/2004SoPh..220..137C} {220, 137}

\bibitem[\protect\citeauthoryear{{Christensen-Dalsgaard}
  et~al.,}{{Christensen-Dalsgaard} et~al.}{2010}]{Christensen2010ApJ}
{Christensen-Dalsgaard} J.,  et~al., 2010, \mn@doi [\apjl]
  {10.1088/2041-8205/713/2/L164}, \href
  {http://adsabs.harvard.edu/abs/2010ApJ...713L.164C} {713, L164}

\bibitem[\protect\citeauthoryear{{Claret} \& {Bloemen}}{{Claret} \&
  {Bloemen}}{2011}]{Claret2011AA}
{Claret} A.,  {Bloemen} S.,  2011, \mn@doi [\aap]
  {10.1051/0004-6361/201116451}, \href
  {http://adsabs.harvard.edu/abs/2011A%26A...529A..75C} {529, A75}

\bibitem[\protect\citeauthoryear{{Czekala}, {Mandel}, {Andrews}, {Dittmann},
  {Ghosh}, {Montet}  \& {Newton}}{{Czekala} et~al.}{2017}]{Czekala2017ApJ}
{Czekala} I.,  {Mandel} K.~S.,  {Andrews} S.~M.,  {Dittmann} J.~A.,  {Ghosh}
  S.~K.,  {Montet} B.~T.,   {Newton} E.~R.,  2017, \mn@doi [\apj]
  {10.3847/1538-4357/aa6aab}, \href
  {http://adsabs.harvard.edu/abs/2017ApJ...840...49C} {840, 49}

\bibitem[\protect\citeauthoryear{{DePoy} et~al.,}{{DePoy} et~al.}{2003}]{CT13}
{DePoy} D.~L.,  et~al., 2003, in {Iye} M.,  {Moorwood} A.~F.~M.,  eds,
  \procspie Vol. 4841, Instrument Design and Performance for Optical/Infrared
  Ground-based Telescopes. pp 827--838, \mn@doi{10.1117/12.459907}

\bibitem[\protect\citeauthoryear{{Dominik} et~al.,}{{Dominik}
  et~al.}{2010}]{MINDSTEp}
{Dominik} M.,  et~al., 2010, \mn@doi [Astronomische Nachrichten]
  {10.1002/asna.201011400}, \href
  {http://adsabs.harvard.edu/abs/2010AN....331..671D} {331, 671}

\bibitem[\protect\citeauthoryear{{Eilers}, {Hogg}, {Rix}  \& {Ness}}{{Eilers}
  et~al.}{2018}]{Eilers2018arXiv}
{Eilers} A.-C.,  {Hogg} D.~W.,  {Rix} H.-W.,   {Ness} M.,  2018, preprint,
  \href {https://ui.adsabs.harvard.edu/#abs/2018arXiv181009466E} {p.
  arXiv:1810.09466} (\mn@eprint {arXiv} {1810.09466})

\bibitem[\protect\citeauthoryear{{Evans}, {Aigrain}, {Gibson}, {Barstow},
  {Amundsen}, {Tremblin}  \& {Mourier}}{{Evans} et~al.}{2015}]{Evans2015MNRAS}
{Evans} T.~M.,  {Aigrain} S.,  {Gibson} N.,  {Barstow} J.~K.,  {Amundsen}
  D.~S.,  {Tremblin} P.,   {Mourier} P.,  2015, \mn@doi [\mnras]
  {10.1093/mnras/stv910}, \href
  {http://adsabs.harvard.edu/abs/2015MNRAS.451..680E} {451, 680}

\bibitem[\protect\citeauthoryear{{Evans} et~al.,}{{Evans} et~al.}{2016}]{Mind}
{Evans} D.~F.,  et~al., 2016, \mn@doi [\aap] {10.1051/0004-6361/201527970},
  \href {http://adsabs.harvard.edu/abs/2016A%26A...589A..58E} {589, A58}

\bibitem[\protect\citeauthoryear{{Foreman-Mackey}, {Hogg}, {Lang}  \&
  {Goodman}}{{Foreman-Mackey} et~al.}{2013}]{Foreman2013PASP}
{Foreman-Mackey} D.,  {Hogg} D.~W.,  {Lang} D.,   {Goodman} J.,  2013, \mn@doi
  [\pasp] {10.1086/670067}, \href
  {http://adsabs.harvard.edu/abs/2013PASP..125..306F} {125, 306}

\bibitem[\protect\citeauthoryear{{Foreman-Mackey}, {Agol}, {Ambikasaran}  \&
  {Angus}}{{Foreman-Mackey} et~al.}{2017}]{Foreman2017AJ}
{Foreman-Mackey} D.,  {Agol} E.,  {Ambikasaran} S.,   {Angus} R.,  2017,
  \mn@doi [\aj] {10.3847/1538-3881/aa9332}, \href
  {http://adsabs.harvard.edu/abs/2017AJ....154..220F} {154, 220}

\bibitem[\protect\citeauthoryear{{Fouqu{\'e}} et~al.,}{{Fouqu{\'e}}
  et~al.}{2010}]{Fouque2010AA}
{Fouqu{\'e}} P.,  et~al., 2010, \mn@doi [\aap] {10.1051/0004-6361/201014053},
  \href {https://ui.adsabs.harvard.edu/abs/2010A&A...518A..51F} {518, A51}

\bibitem[\protect\citeauthoryear{{Gaia Collaboration} et~al.,}{{Gaia
  Collaboration} et~al.}{2016}]{Gaia2016AA}
{Gaia Collaboration} et~al., 2016, \mn@doi [\aap]
  {10.1051/0004-6361/201629272}, \href
  {http://adsabs.harvard.edu/abs/2016A%26A...595A...1G} {595, A1}

\bibitem[\protect\citeauthoryear{{Gaia Collaboration} et~al.,}{{Gaia
  Collaboration} et~al.}{2018a}]{Gaia2018AA}
{Gaia Collaboration} et~al., 2018a, \mn@doi [\aap]
  {10.1051/0004-6361/201833051}, \href
  {http://adsabs.harvard.edu/abs/2018A%26A...616A...1G} {616, A1}

\bibitem[\protect\citeauthoryear{{Gaia Collaboration} et~al.,}{{Gaia
  Collaboration} et~al.}{2018b}]{Gaia2018AA2}
{Gaia Collaboration} et~al., 2018b, \mn@doi [\aap]
  {10.1051/0004-6361/201832916}, \href
  {http://adsabs.harvard.edu/abs/2018A%26A...616A..14G} {616, A14}

\bibitem[\protect\citeauthoryear{{Gibson}, {Aigrain}, {Roberts}, {Evans},
  {Osborne}  \& {Pont}}{{Gibson} et~al.}{2012}]{Gibson2012MNRAS}
{Gibson} N.~P.,  {Aigrain} S.,  {Roberts} S.,  {Evans} T.~M.,  {Osborne} M.,
  {Pont} F.,  2012, \mn@doi [\mnras] {10.1111/j.1365-2966.2011.19915.x}, \href
  {http://adsabs.harvard.edu/abs/2012MNRAS.419.2683G} {419, 2683}

\bibitem[\protect\citeauthoryear{{Gilliland} et~al.,}{{Gilliland}
  et~al.}{2010}]{Gilliand2010PASP}
{Gilliland} R.~L.,  et~al., 2010, \mn@doi [\pasp] {10.1086/650399}, \href
  {http://adsabs.harvard.edu/abs/2010PASP..122..131G} {122, 131}

\bibitem[\protect\citeauthoryear{{Gould}}{{Gould}}{1992}]{Gould1992ApJ}
{Gould} A.,  1992, \mn@doi [\apj] {10.1086/171443}, \href
  {http://adsabs.harvard.edu/abs/1992ApJ...392..442G} {392, 442}

\bibitem[\protect\citeauthoryear{{Gould}}{{Gould}}{1994a}]{Gould1994ApJ2}
{Gould} A.,  1994a, \mn@doi [\apjl] {10.1086/187190}, \href
  {http://adsabs.harvard.edu/abs/1994ApJ...421L..71G} {421, L71}

\bibitem[\protect\citeauthoryear{{Gould}}{{Gould}}{1994b}]{Gould1994ApJ}
{Gould} A.,  1994b, \mn@doi [\apjl] {10.1086/187191}, \href
  {http://adsabs.harvard.edu/abs/1994ApJ...421L..75G} {421, L75}

\bibitem[\protect\citeauthoryear{{Gould}}{{Gould}}{1995a}]{Gould1995ApJ}
{Gould} A.,  1995a, \mn@doi [\apjl] {10.1086/187779}, \href
  {http://adsabs.harvard.edu/abs/1995ApJ...441L..21G} {441, L21}

\bibitem[\protect\citeauthoryear{{Gould}}{{Gould}}{1995b}]{Gould1995ApJ2}
{Gould} A.,  1995b, \mn@doi [\apjl] {10.1086/187933}, \href
  {http://adsabs.harvard.edu/abs/1995ApJ...446L..71G} {446, L71}

\bibitem[\protect\citeauthoryear{{Gould}}{{Gould}}{1997a}]{Hollywood1997}
{Gould} A.,  1997a, in {Ferlet} R.,  {Maillard} J.-P.,   {Raban} B.,  eds,
  Variables Stars and the Astrophysical Returns of the Microlensing Surveys.
  p.~125 (\mn@eprint {} {astro-ph/9608045})

\bibitem[\protect\citeauthoryear{{Gould}}{{Gould}}{1997b}]{Gould1997ApJ}
{Gould} A.,  1997b, \mn@doi [\apj] {10.1086/303942}, \href
  {http://adsabs.harvard.edu/abs/1997ApJ...480..188G} {480, 188}

\bibitem[\protect\citeauthoryear{{Gould}}{{Gould}}{2000}]{Gould2000ApJ}
{Gould} A.,  2000, \mn@doi [\apj] {10.1086/317037}, \href
  {http://adsabs.harvard.edu/abs/2000ApJ...542..785G} {542, 785}

\bibitem[\protect\citeauthoryear{{Gould}}{{Gould}}{2003}]{2003astro.ph.10577G}
{Gould} A.,  2003, arXiv e-prints, \href
  {https://ui.adsabs.harvard.edu/abs/2003astro.ph.10577G} {pp
  astro--ph/0310577}

\bibitem[\protect\citeauthoryear{{Gould}}{{Gould}}{2004}]{Gould2004ApJ}
{Gould} A.,  2004, \mn@doi [\apj] {10.1086/382782}, \href
  {http://adsabs.harvard.edu/abs/2004ApJ...606..319G} {606, 319}

\bibitem[\protect\citeauthoryear{{Gould} \& {Horne}}{{Gould} \&
  {Horne}}{2013}]{Gould2013ApJ}
{Gould} A.,  {Horne} K.,  2013, \mn@doi [\apjl] {10.1088/2041-8205/779/2/L28},
  \href {http://adsabs.harvard.edu/abs/2013ApJ...779L..28G} {779, L28}

\bibitem[\protect\citeauthoryear{{Gould} et~al.,}{{Gould} et~al.}{2010}]{mufun}
{Gould} A.,  et~al., 2010, \mn@doi [\apj] {10.1088/0004-637X/720/2/1073}, \href
  {http://adsabs.harvard.edu/abs/2010ApJ...720.1073G} {720, 1073}

\bibitem[\protect\citeauthoryear{{Gould}, {Huber}, {Penny}  \&
  {Stello}}{{Gould} et~al.}{2015}]{Gould2015JKAS}
{Gould} A.,  {Huber} D.,  {Penny} M.,   {Stello} D.,  2015, \mn@doi [Journal of
  Korean Astronomical Society] {10.5303/JKAS.2015.48.2.093}, \href
  {http://adsabs.harvard.edu/abs/2015JKAS...48...93G} {48, 93}

\bibitem[\protect\citeauthoryear{{Groenewegen}}{{Groenewegen}}{2004}]{Mgiant}
{Groenewegen} M.~A.~T.,  2004, \mn@doi [\mnras]
  {10.1111/j.1365-2966.2004.08121.x}, \href
  {http://adsabs.harvard.edu/abs/2004MNRAS.353..903G} {353, 903}

\bibitem[\protect\citeauthoryear{{Grunblatt}, {Howard}  \&
  {Haywood}}{{Grunblatt} et~al.}{2016a}]{Grunblatt2016IAUFM}
{Grunblatt} S.~K.,  {Howard} A.~W.,   {Haywood} R.~D.,  2016a, \mn@doi [IAU
  Focus Meeting] {10.1017/S1743921316002829}, \href
  {http://adsabs.harvard.edu/abs/2016IAUFM..29A.208G} {29, 208}

\bibitem[\protect\citeauthoryear{{Grunblatt} et~al.,}{{Grunblatt}
  et~al.}{2016b}]{Grunblatt2016AJ}
{Grunblatt} S.~K.,  et~al., 2016b, \mn@doi [\aj] {10.3847/0004-6256/152/6/185},
  \href {http://adsabs.harvard.edu/abs/2016AJ....152..185G} {152, 185}

\bibitem[\protect\citeauthoryear{{Grunblatt} et~al.,}{{Grunblatt}
  et~al.}{2017}]{Grunblatt2017AJ}
{Grunblatt} S.~K.,  et~al., 2017, \mn@doi [\aj] {10.3847/1538-3881/aa932d},
  \href {http://adsabs.harvard.edu/abs/2017AJ....154..254G} {154, 254}

\bibitem[\protect\citeauthoryear{{Harvey}}{{Harvey}}{1985}]{Harvey1985ESASP}
{Harvey} J.,  1985, in {Rolfe} E.,  {Battrick} B.,  eds,  ESA Special
  Publication Vol. 235, Future Missions in Solar, Heliospheric \& Space Plasma
  Physics.

\bibitem[\protect\citeauthoryear{{Hekker} \& {Christensen-Dalsgaard}}{{Hekker}
  \& {Christensen-Dalsgaard}}{2017}]{Hekker2017AApR}
{Hekker} S.,  {Christensen-Dalsgaard} J.,  2017, \mn@doi [\aapr]
  {10.1007/s00159-017-0101-x}, \href
  {http://adsabs.harvard.edu/abs/2017A%26ARv..25....1H} {25, 1}

\bibitem[\protect\citeauthoryear{{Hekker} et~al.,}{{Hekker}
  et~al.}{2012}]{Hekker2012A}
{Hekker} S.,  et~al., 2012, \mn@doi [\aap] {10.1051/0004-6361/201219328}, \href
  {http://adsabs.harvard.edu/abs/2012A%26A...544A..90H} {544, A90}

\bibitem[\protect\citeauthoryear{{Hon}, {Stello}, {Garc{\'{\i}}a}, {Mathur},
  {Sharma}, {Colman}  \& {Bugnet}}{{Hon} et~al.}{2019}]{Hon2019MNRAS}
{Hon} M.,  {Stello} D.,  {Garc{\'{\i}}a} R.~A.,  {Mathur} S.,  {Sharma} S.,
  {Colman} I.~L.,   {Bugnet} L.,  2019, \mn@doi [\mnras]
  {10.1093/mnras/stz622}, \href
  {http://adsabs.harvard.edu/abs/2019MNRAS.tmp..610H} {}

\bibitem[\protect\citeauthoryear{{Huber}, {Stello}, {Bedding}, {Chaplin},
  {Arentoft}, {Quirion}  \& {Kjeldsen}}{{Huber} et~al.}{2009}]{Huber2009CoAst}
{Huber} D.,  {Stello} D.,  {Bedding} T.~R.,  {Chaplin} W.~J.,  {Arentoft} T.,
  {Quirion} P.-O.,   {Kjeldsen} H.,  2009, Communications in Asteroseismology,
  \href {http://adsabs.harvard.edu/abs/2009CoAst.160...74H} {160, 74}

\bibitem[\protect\citeauthoryear{{Huber} et~al.,}{{Huber}
  et~al.}{2010}]{Huber2010CoAst}
{Huber} D.,  et~al., 2010, \mn@doi [\apj] {10.1088/0004-637X/723/2/1607}, \href
  {http://adsabs.harvard.edu/abs/2010ApJ...723.1607H} {723, 1607}

\bibitem[\protect\citeauthoryear{{Huber} et~al.,}{{Huber}
  et~al.}{2011}]{Huber2011ApJ}
{Huber} D.,  et~al., 2011, \mn@doi [\apj] {10.1088/0004-637X/743/2/143}, \href
  {http://adsabs.harvard.edu/abs/2011ApJ...743..143H} {743, 143}

\bibitem[\protect\citeauthoryear{{Huber} et~al.,}{{Huber}
  et~al.}{2013a}]{Huber2013Sci}
{Huber} D.,  et~al., 2013a, \mn@doi [Science] {10.1126/science.1242066}, \href
  {http://adsabs.harvard.edu/abs/2013Sci...342..331H} {342, 331}

\bibitem[\protect\citeauthoryear{{Huber} et~al.,}{{Huber}
  et~al.}{2013b}]{Huber2013ApJ}
{Huber} D.,  et~al., 2013b, \mn@doi [\apj] {10.1088/0004-637X/767/2/127}, \href
  {http://adsabs.harvard.edu/abs/2013ApJ...767..127H} {767, 127}

\bibitem[\protect\citeauthoryear{{Huber} et~al.,}{{Huber}
  et~al.}{2017}]{Huber2017ApJ}
{Huber} D.,  et~al., 2017, \mn@doi [\apj] {10.3847/1538-4357/aa75ca}, \href
  {http://adsabs.harvard.edu/abs/2017ApJ...844..102H} {844, 102}

\bibitem[\protect\citeauthoryear{{Johnson} \& {Yee}}{{Johnson} \&
  {Yee}}{2017}]{Johnson2017PASP}
{Johnson} S.~A.,  {Yee} J.~C.,  2017, \mn@doi [\pasp]
  {10.1088/1538-3873/aa719c}, \href
  {http://adsabs.harvard.edu/abs/2017PASP..129g4401J} {129, 074401}

\bibitem[\protect\citeauthoryear{{Kallinger} et~al.,}{{Kallinger}
  et~al.}{2014}]{Kallinger2014A}
{Kallinger} T.,  et~al., 2014, \mn@doi [\aap] {10.1051/0004-6361/201424313},
  \href {http://adsabs.harvard.edu/abs/2014A%26A...570A..41K} {570, A41}

\bibitem[\protect\citeauthoryear{{Kervella} \& {Fouqu{\'e}}}{{Kervella} \&
  {Fouqu{\'e}}}{2008}]{Kervella2008AA}
{Kervella} P.,  {Fouqu{\'e}} P.,  2008, \mn@doi [\aap]
  {10.1051/0004-6361:200810317}, \href
  {http://adsabs.harvard.edu/abs/2008A%26A...491..855K} {491, 855}

\bibitem[\protect\citeauthoryear{{Kervella}, {Th{\'e}venin}, {Di Folco}  \&
  {S{\'e}gransan}}{{Kervella} et~al.}{2004a}]{Kervella2004AA}
{Kervella} P.,  {Th{\'e}venin} F.,  {Di Folco} E.,   {S{\'e}gransan} D.,
  2004a, \mn@doi [\aap] {10.1051/0004-6361:20035930}, \href
  {http://adsabs.harvard.edu/abs/2004A%26A...426..297K} {426, 297}

\bibitem[\protect\citeauthoryear{{Kervella}, {Bersier}, {Mourard}, {Nardetto},
  {Fouqu{\'e}}  \& {Coud{\'e} du Foresto}}{{Kervella}
  et~al.}{2004b}]{Kervella2004b}
{Kervella} P.,  {Bersier} D.,  {Mourard} D.,  {Nardetto} N.,  {Fouqu{\'e}} P.,
   {Coud{\'e} du Foresto} V.,  2004b, \mn@doi [\aap]
  {10.1051/0004-6361:20041416}, \href
  {http://adsabs.harvard.edu/abs/2004A%26A...428..587K} {428, 587}

\bibitem[\protect\citeauthoryear{{Kim} et~al.,}{{Kim} et~al.}{2016}]{KMT2016}
{Kim} S.-L.,  et~al., 2016, \mn@doi [Journal of Korean Astronomical Society]
  {10.5303/JKAS.2016.49.1.037}, \href
  {http://adsabs.harvard.edu/abs/2016JKAS...49...37K} {49, 37}

\bibitem[\protect\citeauthoryear{{Kim} et~al.,}{{Kim} et~al.}{2018}]{Kim2018a}
{Kim} D.-J.,  et~al., 2018, \mn@doi [\aj] {10.3847/1538-3881/aaa47b}, \href
  {http://adsabs.harvard.edu/abs/2018AJ....155...76K} {155, 76}

\bibitem[\protect\citeauthoryear{{Kjeldsen} \& {Bedding}}{{Kjeldsen} \&
  {Bedding}}{1995}]{Kjeldsen1995AA}
{Kjeldsen} H.,  {Bedding} T.~R.,  1995, \aap, \href
  {http://adsabs.harvard.edu/abs/1995A%26A...293...87K} {293, 87}

\bibitem[\protect\citeauthoryear{{Kjeldsen} et~al.,}{{Kjeldsen}
  et~al.}{2008}]{Kjeldsen2008ApJ}
{Kjeldsen} H.,  et~al., 2008, \mn@doi [\apj] {10.1086/589142}, \href
  {http://adsabs.harvard.edu/abs/2008ApJ...682.1370K} {682, 1370}

\bibitem[\protect\citeauthoryear{{Kjeldsen}, {Bedding}  \&
  {Christensen-Dalsgaard}}{{Kjeldsen} et~al.}{2009}]{Kjeldsen2009IAUS}
{Kjeldsen} H.,  {Bedding} T.~R.,   {Christensen-Dalsgaard} J.,  2009, in {Pont}
  F.,  {Sasselov} D.,   {Holman} M.~J.,  eds,  IAU Symposium Vol. 253,
  Transiting Planets. pp 309--317 (\mn@eprint {arXiv} {0807.0508}),
  \mn@doi{10.1017/S1743921308026537}

\bibitem[\protect\citeauthoryear{{Kjeldsen}, {Christensen-Dalsgaard},
  {Handberg}, {Brown}, {Gilliland}, {Borucki}  \& {Koch}}{{Kjeldsen}
  et~al.}{2010}]{Kjeldsen2010AN}
{Kjeldsen} H.,  {Christensen-Dalsgaard} J.,  {Handberg} R.,  {Brown} T.~M.,
  {Gilliland} R.~L.,  {Borucki} W.~J.,   {Koch} D.,  2010, \mn@doi
  [Astronomische Nachrichten] {10.1002/asna.201011437}, \href
  {http://adsabs.harvard.edu/abs/2010AN....331..966K} {331, 966}

\bibitem[\protect\citeauthoryear{{Kroupa}}{{Kroupa}}{2001}]{Kroupa2001MNRAS}
{Kroupa} P.,  2001, \mn@doi [\mnras] {10.1046/j.1365-8711.2001.04022.x}, \href
  {http://adsabs.harvard.edu/abs/2001MNRAS.322..231K} {322, 231}

\bibitem[\protect\citeauthoryear{{Mathur} et~al.,}{{Mathur}
  et~al.}{2011}]{Mathur2011ApJ}
{Mathur} S.,  et~al., 2011, \mn@doi [\apj] {10.1088/0004-637X/741/2/119}, \href
  {http://adsabs.harvard.edu/abs/2011ApJ...741..119M} {741, 119}

\bibitem[\protect\citeauthoryear{{Michel}, {Samadi}, {Baudin}, {Barban},
  {Appourchaux}  \& {Auvergne}}{{Michel} et~al.}{2009}]{Michel2009A}
{Michel} E.,  {Samadi} R.,  {Baudin} F.,  {Barban} C.,  {Appourchaux} T.,
  {Auvergne} M.,  2009, \mn@doi [\aap] {10.1051/0004-6361:200810353}, \href
  {http://adsabs.harvard.edu/abs/2009A%26A...495..979M} {495, 979}

\bibitem[\protect\citeauthoryear{{Mosser} et~al.,}{{Mosser}
  et~al.}{2012}]{Mosser2011APP}
{Mosser} B.,  et~al., 2012, \mn@doi [\aap] {10.1051/0004-6361/201117352}, \href
  {http://adsabs.harvard.edu/abs/2012A%26A...537A..30M} {537, A30}

\bibitem[\protect\citeauthoryear{{Mosser} et~al.,}{{Mosser}
  et~al.}{2013}]{Mosser2013A}
{Mosser} B.,  et~al., 2013, \mn@doi [\aap] {10.1051/0004-6361/201322243}, \href
  {http://adsabs.harvard.edu/abs/2013A%26A...559A.137M} {559, A137}

\bibitem[\protect\citeauthoryear{{Mr{\'o}z} et~al.,}{{Mr{\'o}z}
  et~al.}{2017}]{Mroz2017AJ}
{Mr{\'o}z} P.,  et~al., 2017, \mn@doi [\aj] {10.3847/1538-3881/aa8f98}, \href
  {http://adsabs.harvard.edu/abs/2017AJ....154..205M} {154, 205}

\bibitem[\protect\citeauthoryear{{Mr{\'o}z} et~al.,}{{Mr{\'o}z}
  et~al.}{2018}]{MrozNeptune}
{Mr{\'o}z} P.,  et~al., 2018, \mn@doi [\aj] {10.3847/1538-3881/aaaae9}, \href
  {http://adsabs.harvard.edu/abs/2018AJ....155..121M} {155, 121}

\bibitem[\protect\citeauthoryear{{Nataf} et~al.,}{{Nataf}
  et~al.}{2016}]{Nataf2016}
{Nataf} D.~M.,  et~al., 2016, \mn@doi [\mnras] {10.1093/mnras/stv2843}, \href
  {http://adsabs.harvard.edu/abs/2016MNRAS.456.2692N} {456, 2692}

\bibitem[\protect\citeauthoryear{{Navarro}, {Minniti}  \&
  {Contreras-Ramos}}{{Navarro} et~al.}{2018}]{Navarro2018ApJ}
{Navarro} M.~G.,  {Minniti} D.,   {Contreras-Ramos} R.,  2018, \mn@doi [\apjl]
  {10.3847/2041-8213/aae08a}, \href
  {http://adsabs.harvard.edu/abs/2018ApJ...865L...5N} {865, L5}

\bibitem[\protect\citeauthoryear{{Nemiroff} \& {Wickramasinghe}}{{Nemiroff} \&
  {Wickramasinghe}}{1994}]{Nemiroff1994ApJ}
{Nemiroff} R.~J.,  {Wickramasinghe} W.~A.~D.~T.,  1994, \mn@doi [\apjl]
  {10.1086/187265}, \href {http://adsabs.harvard.edu/abs/1994ApJ...424L..21N}
  {424, L21}

\bibitem[\protect\citeauthoryear{{Nishiyama}, {Tamura}, {Hatano}, {Kato},
  {Tanab{\'e}}, {Sugitani}  \& {Nagata}}{{Nishiyama} et~al.}{2009}]{N09}
{Nishiyama} S.,  {Tamura} M.,  {Hatano} H.,  {Kato} D.,  {Tanab{\'e}} T.,
  {Sugitani} K.,   {Nagata} T.,  2009, \mn@doi [\apj]
  {10.1088/0004-637X/696/2/1407}, \href
  {http://adsabs.harvard.edu/abs/2009ApJ...696.1407N} {696, 1407}

\bibitem[\protect\citeauthoryear{{OGLE Collaboration} et~al.,}{{OGLE
  Collaboration} et~al.}{2019}]{Mroz2FFP}
{OGLE Collaboration} et~al., 2019, \mn@doi [\aap]
  {10.1051/0004-6361/201834557}, \href
  {https://ui.adsabs.harvard.edu/abs/2019A&A...622A.201O} {622, A201}

\bibitem[\protect\citeauthoryear{{Paczy{\'n}ski}}{{Paczy{\'n}ski}}{1986}]{Paczynski1986ApJ}
{Paczy{\'n}ski} B.,  1986, \mn@doi [\apj] {10.1086/164140}, \href
  {http://adsabs.harvard.edu/abs/1986ApJ...304....1P} {304, 1}

\bibitem[\protect\citeauthoryear{{Penny}, {Henderson}  \& {Clanton}}{{Penny}
  et~al.}{2016}]{Penny2016ApJ}
{Penny} M.~T.,  {Henderson} C.~B.,   {Clanton} C.,  2016, \mn@doi [\apj]
  {10.3847/0004-637X/830/2/150}, \href
  {http://adsabs.harvard.edu/abs/2016ApJ...830..150P} {830, 150}

\bibitem[\protect\citeauthoryear{{Penny}, {Gaudi}, {Kerins}, {Rattenbury},
  {Mao}, {Robin}  \& {Calchi Novati}}{{Penny} et~al.}{2019}]{Penny2018arXiv}
{Penny} M.~T.,  {Gaudi} B.~S.,  {Kerins} E.,  {Rattenbury} N.~J.,  {Mao} S.,
  {Robin} A.~C.,   {Calchi Novati} S.,  2019, \mn@doi [\apjs]
  {10.3847/1538-4365/aafb69}, \href
  {https://ui.adsabs.harvard.edu/abs/2019ApJS..241....3P} {241, 3}

\bibitem[\protect\citeauthoryear{{Ranc} et~al.,}{{Ranc}
  et~al.}{2019}]{Ranc2018arXiv}
{Ranc} C.,  et~al., 2019, \mn@doi [\aj] {10.3847/1538-3881/ab141b}, \href
  {https://ui.adsabs.harvard.edu/abs/2019AJ....157..232R} {157, 232}

\bibitem[\protect\citeauthoryear{{Rasmussen} \& {Williams}}{{Rasmussen} \&
  {Williams}}{2006}]{Rasmussen2006}
{Rasmussen} C.~E.,  {Williams} C.~K.~I.,  2006, {Gaussian Processes for Machine
  Learning}

\bibitem[\protect\citeauthoryear{{Refsdal}}{{Refsdal}}{1966}]{Refsdal1966MNRAS}
{Refsdal} S.,  1966, \mn@doi [\mnras] {10.1093/mnras/134.3.315}, \href
  {http://adsabs.harvard.edu/abs/1966MNRAS.134..315R} {134, 315}

\bibitem[\protect\citeauthoryear{{Ryu} et~al.,}{{Ryu} et~al.}{2018}]{Ryu2018AJ}
{Ryu} Y.~H.,  et~al., 2018, \mn@doi [\aj] {10.3847/1538-3881/aa9be4}, \href
  {https://ui.adsabs.harvard.edu/\#abs/2018AJ....155...40R} {155, 40}

\bibitem[\protect\citeauthoryear{{Saito} et~al.,}{{Saito} et~al.}{2012}]{VVV}
{Saito} R.~K.,  et~al., 2012, \mn@doi [\aap] {10.1051/0004-6361/201118407},
  \href {http://adsabs.harvard.edu/abs/2012A%26A...537A.107S} {537, A107}

\bibitem[\protect\citeauthoryear{{Schechter}, {Mateo}  \& {Saha}}{{Schechter}
  et~al.}{1993}]{dophot}
{Schechter} P.~L.,  {Mateo} M.,   {Saha} A.,  1993, \mn@doi [\pasp]
  {10.1086/133316}, \href {http://adsabs.harvard.edu/abs/1993PASP..105.1342S}
  {105, 1342}

\bibitem[\protect\citeauthoryear{{Sch{\"o}nrich}, {Binney}  \&
  {Dehnen}}{{Sch{\"o}nrich} et~al.}{2010}]{Schonrich2010MNRAS}
{Sch{\"o}nrich} R.,  {Binney} J.,   {Dehnen} W.,  2010, \mn@doi [\mnras]
  {10.1111/j.1365-2966.2010.16253.x}, \href
  {http://adsabs.harvard.edu/abs/2010MNRAS.403.1829S} {403, 1829}

\bibitem[\protect\citeauthoryear{{Shin} et~al.,}{{Shin}
  et~al.}{2018}]{Shin2018ApJ}
{Shin} I.-G.,  et~al., 2018, \mn@doi [\apj] {10.3847/1538-4357/aacdf4}, \href
  {http://ads.bao.ac.cn/abs/2018ApJ...863...23S} {863, 23}

\bibitem[\protect\citeauthoryear{{Shvartzvald}, {Bryden}, {Gould}, {Henderson},
  {Howell}  \& {Beichman}}{{Shvartzvald} et~al.}{2017}]{Shvartzvald2017AJ}
{Shvartzvald} Y.,  {Bryden} G.,  {Gould} A.,  {Henderson} C.~B.,  {Howell}
  S.~B.,   {Beichman} C.,  2017, \mn@doi [\aj] {10.3847/1538-3881/153/2/61},
  \href {http://adsabs.harvard.edu/abs/2017AJ....153...61S} {153, 61}

\bibitem[\protect\citeauthoryear{{Shvartzvald} et~al.,}{{Shvartzvald}
  et~al.}{2018}]{Shvartzvald2018ApJ}
{Shvartzvald} Y.,  et~al., 2018, \mn@doi [\apjl] {10.3847/2041-8213/aab71b},
  \href {http://adsabs.harvard.edu/abs/2018ApJ...857L...8S} {857, L8}

\bibitem[\protect\citeauthoryear{{Shvartzvald} et~al.,}{{Shvartzvald}
  et~al.}{2019}]{Shvartzvald2018arXiv}
{Shvartzvald} Y.,  et~al., 2019, \mn@doi [\aj] {10.3847/1538-3881/aafe12},
  \href {https://ui.adsabs.harvard.edu/abs/2019AJ....157..106S} {157, 106}

\bibitem[\protect\citeauthoryear{{Skowron} et~al.,}{{Skowron}
  et~al.}{2016}]{Skowron2016AcA}
{Skowron} J.,  et~al., 2016, \actaa, \href
  {http://adsabs.harvard.edu/abs/2016AcA....66....1S} {66, 1}

\bibitem[\protect\citeauthoryear{{Soszy{\'n}ski} et~al.,}{{Soszy{\'n}ski}
  et~al.}{2013}]{Soszynski2013AcA}
{Soszy{\'n}ski} I.,  et~al., 2013, \actaa, \href
  {http://adsabs.harvard.edu/abs/2013AcA....63...21S} {63, 21}

\bibitem[\protect\citeauthoryear{{Spergel} et~al.,}{{Spergel}
  et~al.}{2015}]{Spergel2015arXiv}
{Spergel} D.,  et~al., 2015, preprint, \href
  {http://adsabs.harvard.edu/abs/2015arXiv150303757S} {} (\mn@eprint {arXiv}
  {1503.03757})

\bibitem[\protect\citeauthoryear{{Stello} et~al.,}{{Stello}
  et~al.}{2009}]{Stello2009ApJ}
{Stello} D.,  et~al., 2009, \mn@doi [\apj] {10.1088/0004-637X/700/2/1589},
  \href {http://adsabs.harvard.edu/abs/2009ApJ...700.1589S} {700, 1589}

\bibitem[\protect\citeauthoryear{{Stello} et~al.,}{{Stello}
  et~al.}{2014}]{Stello2014ApJL}
{Stello} D.,  et~al., 2014, \mn@doi [\apjl] {10.1088/2041-8205/788/1/L10},
  \href {http://adsabs.harvard.edu/abs/2014ApJ...788L..10S} {788, L10}

\bibitem[\protect\citeauthoryear{{Street} et~al.,}{{Street}
  et~al.}{2013}]{Street2013ApJ}
{Street} R.~A.,  et~al., 2013, \mn@doi [\apj] {10.1088/0004-637X/763/1/67},
  \href {http://adsabs.harvard.edu/abs/2013ApJ...763...67S} {763, 67}

\bibitem[\protect\citeauthoryear{{Street} et~al.,}{{Street}
  et~al.}{2016}]{Street2016ApJ}
{Street} R.~A.,  et~al., 2016, \mn@doi [\apj] {10.3847/0004-637X/819/2/93},
  \href {http://adsabs.harvard.edu/abs/2016ApJ...819...93S} {819, 93}

\bibitem[\protect\citeauthoryear{{Sumi} et~al.,}{{Sumi} et~al.}{2016}]{MOA2016}
{Sumi} T.,  et~al., 2016, \mn@doi [\apj] {10.3847/0004-637X/825/2/112}, \href
  {http://adsabs.harvard.edu/abs/2016ApJ...825..112S} {825, 112}

\bibitem[\protect\citeauthoryear{{Udalski}, {Szymanski}, {Soszynski}  \&
  {Poleski}}{{Udalski} et~al.}{2008}]{Udalski2008AcA}
{Udalski} A.,  {Szymanski} M.~K.,  {Soszynski} I.,   {Poleski} R.,  2008,
  \actaa, \href {http://adsabs.harvard.edu/abs/2008AcA....58...69U} {58, 69}

\bibitem[\protect\citeauthoryear{{Udalski}, {Szyma{\'n}ski}  \&
  {Szyma{\'n}ski}}{{Udalski} et~al.}{2015a}]{OGLEIV}
{Udalski} A.,  {Szyma{\'n}ski} M.~K.,   {Szyma{\'n}ski} G.,  2015a, \actaa,
  \href {http://adsabs.harvard.edu/abs/2015AcA....65....1U} {65, 1}

\bibitem[\protect\citeauthoryear{{Udalski} et~al.,}{{Udalski}
  et~al.}{2015b}]{Udalski2015ApJ}
{Udalski} A.,  et~al., 2015b, \mn@doi [\apj] {10.1088/0004-637X/799/2/237},
  \href {http://adsabs.harvard.edu/abs/2015ApJ...799..237U} {799, 237}

\bibitem[\protect\citeauthoryear{{Ulrich}}{{Ulrich}}{1986}]{Ulrich1986ApJ}
{Ulrich} R.~K.,  1986, \mn@doi [\apjl] {10.1086/184700}, \href
  {http://adsabs.harvard.edu/abs/1986ApJ...306L..37U} {306, L37}

\bibitem[\protect\citeauthoryear{{Wang} et~al.,}{{Wang} et~al.}{2017}]{MB15020}
{Wang} T.,  et~al., 2017, \mn@doi [\apj] {10.3847/1538-4357/aa813b}, \href
  {http://adsabs.harvard.edu/abs/2017ApJ...845..129W} {845, 129}

\bibitem[\protect\citeauthoryear{{Wang} et~al.,}{{Wang}
  et~al.}{2018}]{Wang2018ApJ}
{Wang} T.,  et~al., 2018, \mn@doi [\apj] {10.3847/1538-4357/aabcd2}, \href
  {http://adsabs.harvard.edu/abs/2018ApJ...860...25W} {860, 25}

\bibitem[\protect\citeauthoryear{{Witt} \& {Mao}}{{Witt} \&
  {Mao}}{1994}]{Witt1994ApJ}
{Witt} H.~J.,  {Mao} S.,  1994, \mn@doi [\apj] {10.1086/174426}, \href
  {http://adsabs.harvard.edu/abs/1994ApJ...430..505W} {430, 505}

\bibitem[\protect\citeauthoryear{{Wozniak}}{{Wozniak}}{2000}]{Wozniak2000}
{Wozniak} P.~R.,  2000, \actaa, \href
  {http://adsabs.harvard.edu/abs/2000AcA....50..421W} {50, 421}

\bibitem[\protect\citeauthoryear{{Yee} et~al.,}{{Yee}
  et~al.}{2012}]{Yee2012ApJ}
{Yee} J.~C.,  et~al., 2012, \mn@doi [\apj] {10.1088/0004-637X/755/2/102}, \href
  {http://ads.bao.ac.cn/abs/2012ApJ...755..102Y} {755, 102}

\bibitem[\protect\citeauthoryear{{Yee} et~al.,}{{Yee}
  et~al.}{2015a}]{Yee2015ApJ}
{Yee} J.~C.,  et~al., 2015a, \mn@doi [\apj] {10.1088/0004-637X/802/2/76}, \href
  {http://adsabs.harvard.edu/abs/2015ApJ...802...76Y} {802, 76}

\bibitem[\protect\citeauthoryear{{Yee} et~al.,}{{Yee}
  et~al.}{2015b}]{Yee2015ApJ.criteria}
{Yee} J.~C.,  et~al., 2015b, \mn@doi [\apj] {10.1088/0004-637X/810/2/155},
  \href {http://adsabs.harvard.edu/abs/2015ApJ...810..155Y} {810, 155}

\bibitem[\protect\citeauthoryear{{Yoo} et~al.,}{{Yoo}
  et~al.}{2004}]{Yoo2004ApJ}
{Yoo} J.,  et~al., 2004, \mn@doi [\apj] {10.1086/381241}, \href
  {http://adsabs.harvard.edu/abs/2004ApJ...603..139Y} {603, 139}

\bibitem[\protect\citeauthoryear{{Yu}, {Huber}, {Bedding}, {Stello}, {Hon},
  {Murphy}  \& {Khanna}}{{Yu} et~al.}{2018}]{Yu2018ApJS}
{Yu} J.,  {Huber} D.,  {Bedding} T.~R.,  {Stello} D.,  {Hon} M.,  {Murphy}
  S.~J.,   {Khanna} S.,  2018, \mn@doi [\apjs] {10.3847/1538-4365/aaaf74},
  \href {http://adsabs.harvard.edu/abs/2018ApJS..236...42Y} {236, 42}

\bibitem[\protect\citeauthoryear{{Zhu} et~al.,}{{Zhu}
  et~al.}{2015}]{Zhu2015ApJ}
{Zhu} W.,  et~al., 2015, \mn@doi [\apj] {10.1088/0004-637X/805/1/8}, \href
  {http://adsabs.harvard.edu/abs/2015ApJ...805....8Z} {805, 8}

\bibitem[\protect\citeauthoryear{{Zhu} et~al.,}{{Zhu} et~al.}{2017}]{Zhu2017AJ}
{Zhu} W.,  et~al., 2017, \mn@doi [\aj] {10.3847/1538-3881/aa8ef1}, \href
  {http://adsabs.harvard.edu/abs/2017AJ....154..210Z} {154, 210}

\makeatother
\end{thebibliography}




\bsp	
\label{lastpage}
\end{document}